\documentclass[twocolumn,numberedappendix,twocolappendix,appendixfloats]{openjournal_dhayaa}

\usepackage{natbib}
\setcitestyle{aysep={}}
\usepackage{pifont}

\usepackage[colorlinks=true
  ,urlcolor=blue
  ,anchorcolor=blue
  ,citecolor=blue
  ,filecolor=blue
  ,linkcolor=blue
  ,menucolor=blue
  ,linktocpage=true
  ,pdfproducer=medialab
  ,pdfa=true
]{hyperref}

\usepackage{xcolor}

\definecolor{amaranth}{rgb}{0.9, 0.17, 0.31}

\usepackage{graphicx}	
\usepackage{amsmath}	

\usepackage[T1]{fontenc}
\usepackage{ae,aecompl}

\newcommand{\OrcidID}[1]{ \href[urlcolor = red]{https://orcid.org/#1}{\textcolor{lightgray}{\faOrcid}}}
\newcommand{\OrcidIDName}[2]{\href{https://orcid.org/#1}{#2}}

\newcommand{\code}[1]{\texttt{#1}}
\newcommand{\fnl}{f_{\text{NL}}}
\newcommand{\jax}{\code{jax\_cosmo}}
\newcommand{\kpeak}{k_{\mathrm{peak}}}
\newcommand{\same}{\textquotesingle\textquotesingle}

\begin{document}

\title[Cosmology on the largest scales with galaxies and gravitational waves]
{Cosmological Constraints from Combining Photometric Galaxy Surveys and Gravitational Wave Observatories}

\author{\OrcidIDName{0000-0002-9970-3034}{E.~L.~Gagnon}\altaffilmark{*, 1, 2}, \OrcidIDName{0000-0003-3312-909X}{D.~Anbajagane}\altaffilmark{1, 2}, \OrcidIDName{0000-0002-5933-5150}{J.~Prat}\altaffilmark{1, 2}, 
\OrcidIDName{0000-0002-7887-0896}{C.~Chang}\altaffilmark{1, 2}, 
\OrcidIDName{0000-0003-4079-3263}{J.~Frieman}\altaffilmark{1, 2, 3}\\}

\affiliation{$^{1}$ Department of Astronomy and Astrophysics, University of Chicago, Chicago, IL 60637, USA}
\affiliation{$^{2}$ Kavli Institute for Cosmological Physics, University of Chicago, Chicago, IL 60637, USA}
\affiliation{$^{3}$ Fermi National Accelerator Laboratory, Batavia, IL 60510, USA\\}

\email[$^\ast$: Corresponding author: ]{elgagnon@uchicago.edu}


\label{firstpage}

\begin{abstract}
Spatial variations in survey properties due to both observational and astrophysical selection effects generate substantial systematic errors in large-scale structure measurements in optical galaxy surveys on very large scales. On such scales, the statistical sensitivity of optical surveys is also limited by their finite sky coverage. By contrast, gravitational wave (GW) sources appear to be relatively free of these issues, provided the angular sensitivity of GW experiments can be accurately characterized. We quantify the expected cosmological information gain from combining the forecast LSST 3$\times$2pt analysis (the combination of three two-point correlations of galaxy density and weak lensing shear fields) with the large-scale auto-correlation of GW sources from proposed next-generation GW experiments. We find that in $\Lambda$CDM and $w$CDM models, there is no significant improvement in cosmological constraints from combining GW with LSST 3$\times$2pt over LSST alone, due to the large shot noise for the former; however, this combination does enable an estimated $\sim6\%$ constraint on the linear galaxy bias of GW sources. More interestingly, the optical-GW data combination provides tight constraints on models with primordial non-Gaussianity (PNG), due to the predicted scale-dependent bias in PNG models on large scales. Assuming that the largest angular scales that LSST will probe are comparable to those in Stage III surveys ($\ell_{\rm min} \sim 50$), the inclusion of next-generation GW measurements could improve constraints on the PNG parameter $\fnl$ by up to a factor of $\simeq 6.6$ compared to LSST alone, yielding $\sigma(\fnl)=8.5$. These results assume the expected capability of a network of Einstein Telescope-like GW observatories, with a GW detection rate of $10^6$ events/year. We investigate the sensitivity of our results to different assumptions about future GW detectors as well as different LSST analysis choices.

\end{abstract}



\section{Introduction}

As cosmic surveys continue to grow in scale, they enable measurements of large-scale structure (LSS) with ever-greater precision that in principle translate into ever-tighter cosmological constraints. The current state of the art in extracting cosmological information from large photometric surveys is the so-called ``3$\times$2pt'' analysis, which combines the information from three two-point correlation functions: the galaxy auto-correlation function, the weak lensing (cosmic shear, or shear-shear) correlation function, and the galaxy-shear correlation function. All three major Stage III photometric survey collaborations\footnote{The Stage-III and Stage-IV classification was introduced in the Dark Energy Task Force report \citep{Albrecht2006}, where Stage-III refers to the dark energy experiments that started data taking in the 2010s and Stage-IV to those that started or will start in the 2020s.} have carried out such analyses: the Dark Energy Survey (DES, \citealt{desy1-3x2, desy3-3x2}), the Kilo-Degree Survey (KiDS, \citealt{kids1000-3x2}) and the Hyper Suprime-Cam Subaru Strategic Program (HSC-SPP, \citealt{hsc-3x2}). 

While these correlations can be measured over a very wide range of length scales, so far cosmological inference from them has focused on intermediate spatial scales only -- typically between $\sim 8 h^{-1}$ Mpc and $\sim 100 h^{-1}$Mpc -- due to limitations at both larger and smaller scales. Smaller scale measurements, although having higher signal-to-noise, are challenging to model due to nonlinearities in the density field, uncertainties in the galaxy-dark matter connection, and complex baryon physics \citep{Wechsler2018, Martinelli, Chen2023}. Structure on very large scales is easier to model, but the measurements suffer from observational and astrophysical systematic effects. For example, in the Year 3 (Y3) DES galaxy clustering measurements, it was shown that significant corrections (larger than 5$\sigma$, where $\sigma$ is the measurement uncertainty) had to be made to the measurements due to observational systematic effects on angular scales larger than 250 arcmin ($\ell \lesssim 45$), as shown in Fig. 2 of \citet{desy3-clustering}. To fully exploit the cosmological information coming from galaxy surveys, it is imperative to re-examine some of these scale limitations with new tools and data that become available over time.

While there has been substantial recent work employing emerging techniques and simulations to harness information from small scales, e.g., \citet{Miyatake2022, Kobayashi2022, Yuan2022, hsc-3x2-small, Arico2023, Dvornik2023, Lange2023}, this paper focuses on extracting information from the largest scales. The Vera C. Rubin Observatory's Legacy Survey of Space and Time  \citep[LSST,][]{Ivezic2019} will enable a 3$\times$2pt analysis on large scales of unprecedented statistical power. But as an optical photometric survey, it will be subject to the same sources of astrophysical and observational systematics as Stage III surveys. Therefore, we consider using gravitational wave (GW) sources as a complementary probe of very large-scale structure and combining GW and LSST measurements.

The rationale of this approach is that on very large scales, GW observations suffer significantly less from observational and astrophysical systematic effects compared to optically selected galaxies, and their selection function can be known extremely well \citep{Schutz2011, Chen2017} -- unlike electromagnetic signals, GWs are not affected by issues such as spatially varying Galactic dust extinction, variations in survey depth due to time-varying observing conditions and spatially varying star density, and other similar phenomena. In addition, gravitational waveforms are a direct prediction of general relativity \citep[eg.][]{Finn1993}.
Moreover, a distinctive feature of GW observations is their full sky coverage, a significant advantage over ground-based electromagnetic observations, which typically have limited sky coverage due to their geographic location and the need to discard a substantial portion of the sky ($25-40\%$) to obtain cosmological information, due to Galactic foregrounds. GW sources can thus be used to robustly measure clustering on very large angular scales.

With a limited number of terrestrial GW detectors simultaneously operating, most GW sources cannot be localized on the sky with an angular precision better than a few square degrees. 
As a result, the angular clustering of GW sources cannot be measured on small scales. This is not a limitation for our analysis, however, as LSST will provide high signal-to-noise measurements on such scales. 

As we will see, combining LSST $3\times2$pt with GW clustering also allows us to break the degeneracy between the bias of GW host-galaxy sources, $b_{\rm GW}$, and the amplitude of mass clustering in the universe, $\sigma_8$, yielding interesting constraints on $b_{\rm GW}$. This is of interest in its own right, since the bias of GW host-galaxies is not currently well determined (e.g. \citealt{Adhikari2020, Zheng2023}), and constraining it will have implications for understanding the formation channel of GW sources \citep{Scelfo2018}.

Previous works have studied the clustering of GW sources and its cross-correlation with galaxy positions. \citet{Oguri2016, Bosi2023,Scelfo2023} considered GW$\times$LSS cross-correlations to constrain cosmological models, including modified gravity theories. \citet{Bera_2020, Cigarran2022}  used cross-correlations to infer $H_0$, and \citet{Mukherjee_2021} constrained cosmological parameters including the GW bias.  Other studies have focused on astrophysical understanding of GW sources \citep{Scelfo2018, Scelfo2020, Calore2020}, realistic observational effects \citep{Shao2022, Yang2023}, weak lensing of GW events \citep{Balaudo2023, Mpetha2023}, and cross-correlations with other tracers like supernovae \citep{Libanore2022} and HI intensity mapping \citep{Scelfo2022}. Synergies between galaxy surveys and GW observations at the catalog level have also been explored to understand better the binary-host connection e.g. \citet{Adhikari2020} and to help infer both cosmological and astrophysical information from dark sirens e.g. \citet{Gray2023, Borghi2024}.

Our paper takes a different direction from previous work in two ways: 1) we study the complementary combination of GW clustering and LSST 3$\times$2pt, and 2) we explore constraints on both the standard $\Lambda$CDM and $w$CDM cosmological models as well as primordial non-Gaussianity (PNG), specifically, the local PNG model parametrized by $\fnl$. To date, the \textit{Planck} cosmic microwave background (CMB) measurements have provided the most stringent constraints on PNG \citep{Planck-fnl}, but these measurements are already near the cosmic variance limit of precision, so future CMB experiments are unlikely to significantly improve upon these results. Thus, the new frontier for constraining PNG lies in large-scale surveys of the galaxy and matter distributions. We note that \citet{Namikawa2016} also considered GW clustering for constraining $f_{\mathrm{NL}}$, but they did not combine GW with galaxy $3\times2$pt analyses, which offers some major advantages that we explore in this work, such as the self-calibration of the GW host-galaxy bias.

The LSST analysis choices have already been thoroughly studied \citep{DESCSRD} and have been set using information from surveys of the current decade. In our work, we fix the LSST analysis to these choices, and study a wide variety of choices for the GW experiment/sample. This is done to understand the type of GW experiment that can best complements the expected measurements from LSST. Given this goal to study a wide variety of GW setups, we consider many best-case scenario models and some futuristic models as well; in all cases we highlight this during the discussions of the same.

This paper is structured as follows. In Sec.~\ref{sec:modeling}, we present the theoretical framework, in Sec.~\ref{sec:setup}, we describe the analysis setups for galaxy surveys and GWs, and in Sec.~\ref{sec:results} we present the results of our analysis. We conclude in Sec.~\ref{sec:conclusions}.


\section{Modeling} \label{sec:modeling}

\subsection{Large-Scale Structure in $\Lambda$CDM and $w$CDM}

Given the limited photometric redshift precision achieved by imaging surveys that employ a handful of optical passbands, the lowest order LSS clustering statistic commonly measured is the 2D angular correlation function or equivalently the 2D angular power spectrum, $C^{ij}(\ell)$, within and between photometric-redshift bins $i,j$. Here, $\ell$ is the 2D multipole moment, which is roughly related to angular separation on the sky, $\theta$, by $\ell\sim\pi/\theta$. In the $3\times2$pt analysis, we focus on three angular power spectra: the auto-correlation of a foreground (or lens) galaxy population, $C_{\rm gg}$, the cross-correlation between lens-galaxy position and source-galaxy shear, $C_{\rm g \gamma}$, also known as galaxy-galaxy lensing, and the auto-correlation of source-galaxy shear, $C_{\gamma \gamma}$, also known as cosmic shear.

We follow common practice in using the 
first-order Limber approximation \citep{limber} to relate the angular galaxy-galaxy, galaxy-shear, and shear-shear power spectra to the corresponding 3D power spectra, 

\begin{subequations}
\begin{equation}
    C_{\rm gg}^{ij}(\ell) = \int d\chi\frac{N_l^i(\chi)N_{l}^j(\chi)}{\chi^2}P_{\rm gg}\bigg{(}k=\frac{\ell+1/2}{\chi}, z(\chi)\bigg{)},
\end{equation}

\begin{equation}
    C_{\rm g \gamma}^{ij}(\ell) = \int d\chi \frac{N_{l}^i(\chi)q_s^j(\chi)}{\chi^2}P_{\rm gm}\bigg{(}k = \frac{\ell+1/2}{\chi}, z(\chi)\bigg{)},
\end{equation}
and
\begin{equation}
    C_{\gamma\gamma}^{ij}(\ell) = \int d \chi \frac{q^i_s(\chi)q_s^j(\chi)}{\chi^2} P_{\rm mm} \bigg{(} k = \frac{\ell + 1/2}{\chi}, z(\chi) \bigg{)}.
\end{equation}
\end{subequations}
Here, $P_{\rm mm}(k,z)$ is the 3D matter power spectrum, $P_{\rm gm}(k,z)$ is the 3D (lens) galaxy-matter power spectrum, and $P_{\rm gg}(k,z)$ is the 3D lens-galaxy power spectrum, $\chi$ is the comoving distance, $z$ is redshift, and $N^i_l$ describes the true redshift distribution of the lens-galaxy population in the $i$th photometric-redshift bin:
\begin{equation}
N_l^i(\chi) = \frac{n^i_l\,(z)}{\bar{n}^i_l}\frac{dz}{d\chi}, 
\end{equation}
where $n^i_l$ is the lens-galaxy redshift distribution, and $\bar{n}^i_l$ is the mean number density of lens galaxies. In addition, $q_s^{i}$ is the lensing kernel of the source-galaxy population:
\begin{equation} \label{eq:lensing_window}
q_s^i(\chi) = \frac{3H^2_0 \Omega_m }{2c^2} p(\ell) \frac{\chi}{a(\chi)} g^i(\chi),
\end{equation}
where $a$ is the scale factor, $p(\ell)$ is the $\ell$-dependent prefactor in the lensing observables due to the spin-2 nature\footnote{Here $p(\ell) = \sqrt{(\ell - 1)(\ell)(\ell + 1)(\ell + 2)}/(\ell + 0.5)^2$; see Equations (28) and (73) in \citet{Chisari:2019:CCL}}
of the shear, and $g(\chi)$ is the lensing efficiency kernel:
\begin{equation}
 g^{i}(\chi) = \int_\chi^{\chi_\text{h}} d \chi'  \frac{n^i_s\,(z) }{\bar{n}^i_s} \frac{dz}{d\chi'}\frac{\chi'- \chi}{\chi'},
\end{equation}
where $n_s^i(z)$ is the redshift distribution of the source galaxies in the $i$th photometric redshift bin, $\bar{n}^i_s$ is the mean number density of the source galaxies, and $\chi_\text{h}$ is the comoving distance to the horizon. 

The Limber approximation assumed in Eqn.(1) is known to break down at large angular scales, that is, for small $\ell$. This is particularly the case if the kernel functions $N^i_l(\chi)$ and $q^i_s(\chi)$ are narrowly peaked functions of $\chi$. Given the relatively broad redshift bins we adopt in this analysis (see Fig. \ref{fig:n(z)s}) and scaling from the results in \cite{limber}, we expect this to lead to a relatively small error in our analysis for angular multipoles $\ell \gtrsim 3$, one outweighed by the significant savings in computation time. As an additional complication, the mean of any estimator of $C_\ell$ in a survey of finite angular extent $\Theta$ will differ appreciably from the theoretical values in Eqn. (1) for multipoles $\ell < \pi/\Theta$. In practice, for the analysis set-up we consider, this is a very small correction, since GW surveys are effectively all-sky, and $\pi/\Theta_{\rm LSST}$ is comfortably below the minimum angular multipole that we consider in the analysis of LSST data.

Throughout the paper, if not otherwise specified, we take the matter power spectrum $P_{\rm mm}$ to be that given by the spatially flat 5-parameter $\Lambda$CDM model with fiducial parameters $\Omega_c = 0.22$, $\sigma_8=0.8$, $\Omega_b = 0.0448$, $h=0.71$, and $n_s=0.963$. We also consider the $w$CDM model, with an additional free parameter given by the (constant) equation of state parameter of dark energy, $w_0$, with fiducial value $w_0=-1$. The non-linear regime of the matter power spectrum is modeled using the \textsc{Halofit} prescription \citep{Takahashi2012Halofit}. Since we are working on large length scales, we assume a linear, scale-independent galaxy bias model for the lens galaxies, in which 
\begin{equation}
P_{\rm gg}(k,z) = b_{\rm g}(z) P_{\rm gm}(k,z)=b_{\rm g}^2(z)P_{\rm mm}(k,z)~. 
\end{equation}

For GW sources, the populations of lens galaxies in Eqns. (1a) and (2) are replaced by those of GW sources, and we assume that the host galaxies of GW sources are also linearly biased with respect to the matter distribution but with a different bias parameter, $b_{\rm GW}(z)$. Assuming that GW sources can be typically localized to within an area of size $\Theta^2_{\rm GW}$, we only consider clustering of GW sources on angular scales $\ell < \pi/\Theta_{\rm GW}$, so that localization corrections to the estimator for Eqn. (1a) should be very small.

To calculate angular power spectra and their covariances (see \S 3), 
we use the new differentiable cosmology library $\jax$ \citep{jax_paper} to perform our analysis. Testing a variety of setups (see Sec.~\ref{sec:LSST setup},~\ref{sec:gw setup}) is core to our investigation, and $\jax$'s differentiable functions provide the crucial ability to rapidly perform Fisher forecasts in a robust and stable way.

\subsection{Primordial Non-Gaussianity and \textit{f}$_{\text{NL}}$}\label{sec:fnl_into}

In addition to the canonical $\Lambda$CDM and $w$CDM models, we consider $\Lambda$CDM models with non-Gaussian initial conditions. Inflation is an epoch of rapid expansion in the early Universe that is thought to generate the primordial density fluctuations that subsequently evolve into the structures observed today. In the simplest inflation models involving a single, weakly coupled scalar field, the initial density perturbations are predicted to be very close to Gaussian random fields. In models with multiple scalar fields, the initial perturbations can be (potentially measurably) non-Gaussian. 

We focus on models with local-type primordial non-Gaussianity (PNG) \citep{Byrnes2010fNLLoc}, in which the initial conditions for the gravitational potential are given by
\begin{equation}
    \Phi(\vec{\boldsymbol{x}}) = \phi(\vec{\boldsymbol{x}}) + \fnl(\phi(\vec{\boldsymbol{x}})^2 - \langle\phi(\vec{\boldsymbol{x}})^2\rangle),
\end{equation}
where $\Phi(\vec{\boldsymbol{x}})$ is the non-gaussian gravitational potential, $\phi(\vec{\boldsymbol{x}})$ is the original, gaussian potential, $\langle\phi(\vec{\boldsymbol{x}})^2\rangle$ is the average of the squared field, and $\fnl$ is a parameter that characterizes the amplitude of the PNG. In the simplest single-field inflation models, we generically expect $\fnl \ll 1$, so any indication of $\fnl>1$ would imply that multiple fields were present during inflation \citep{Byrnes2010fNLLoc, InflationTheoryObs}.

PNG alters the behavior of the galaxy bias and causes galaxies to exhibit a scale-dependent bias with an amplitude that grows on large scales \citep{Dalal2008fNL}; to the lowest order, the galaxy bias in Eqn. (5) is replaced by 

\begin{equation}
   b_{\rm g}(z) \rightarrow b_{\rm g}(z)+b_{\fnl}(k,z)
\label{eqn:fnl}
\end{equation}
where the additional scale-dependent bias term is given by
\begin{equation}\label{eqn:b_fnl}
    b_{\fnl}(k,z) = 3\delta_c(b_{\rm g}(z) - 1)\frac{\fnl}{k^2}\bigg(\frac{\Omega_mH_0^2}{c^2D(z)T(k)}\bigg)~.
\end{equation}
Here, $\delta_c = 1.686$ is the critical threshold for halo collapse in the spherical collapse model, $D(z)$ is the linear growth factor (relative to the present), and $T(z)$ is the $\Lambda$CDM transfer function. The additional bias term grows towards large-scales, as it is proportional to $b_{\fnl} \propto 1/k^2$
\citep{Dalal2008fNL}. 
Formally, the choice above of $(b_{\rm g} - 1)$ can be generalized to $(b_{\rm g} - p)$, where $p$ is some constant. Most works employ $p = 1$, which is derived by assuming universality of the halo mass function. However, there are motivations for selecting other choices for $p$ as well \citep{Barreira2020}. We employ $p = 1$ to be consistent with previous works. Alternative choices of $p$ that are higher (lower) than this value will result in weaker (stronger) $\fnl$ constraints, since lower values of $p$ increase the scale-dependent bias, $b_{\fnl}$, for galaxies with $b_{\rm g} >1$.

To leading order, $\fnl$ impacts structure formation only through the bias terms in $P_{\rm gg}$ and $P_{\rm gm}$, and we do not include its impact on the matter power spectrum itself. This is a good approximation for linear/quasi-linear scales, as the $\fnl$-based correction goes as $\Delta P_{\rm mm}/P_{\rm mm} \sim \fnl \times 10^{-5}$. For values of $\fnl \sim \mathcal{O}(10)$, as found in our constraints below, this correction is completely subdominant. However, simulations show that the impact of $\fnl$ on the matter power spectrum is most prominent on non-linear scales \citep{Coulton2023PNG, Anbajagane2023}, since the non-Gaussianity of the initial density field changes the abundance of massive halos, which impacts the matter and halo power spectrum on small scales. In principle, this means that our analysis underestimates the full impact of $\fnl$ and thus overestimates the expected error on it. \citet[see their Table 3]{Anbajagane2023} show that a lensing-only analysis of LSST Year 10, with realistic scale cuts, leads to constraints of $\sigma(\fnl) = 92$. This is broader than the constraints on $\fnl$ found in our analysis below, which indicates that the inclusion of $\fnl$ in the matter power spectrum would have negligible impact on our constraints compared to the signal corresponding to scale-dependent bias. 

\begin{figure*}
\begin{center}
\includegraphics[width = 0.32\textwidth]{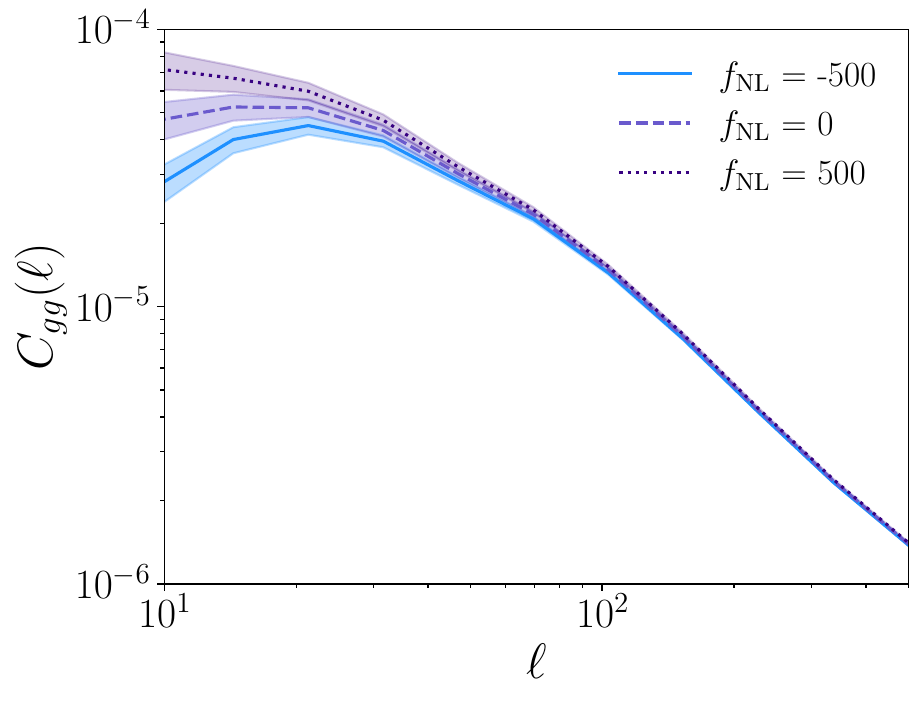}
\includegraphics[width = 0.32\textwidth]{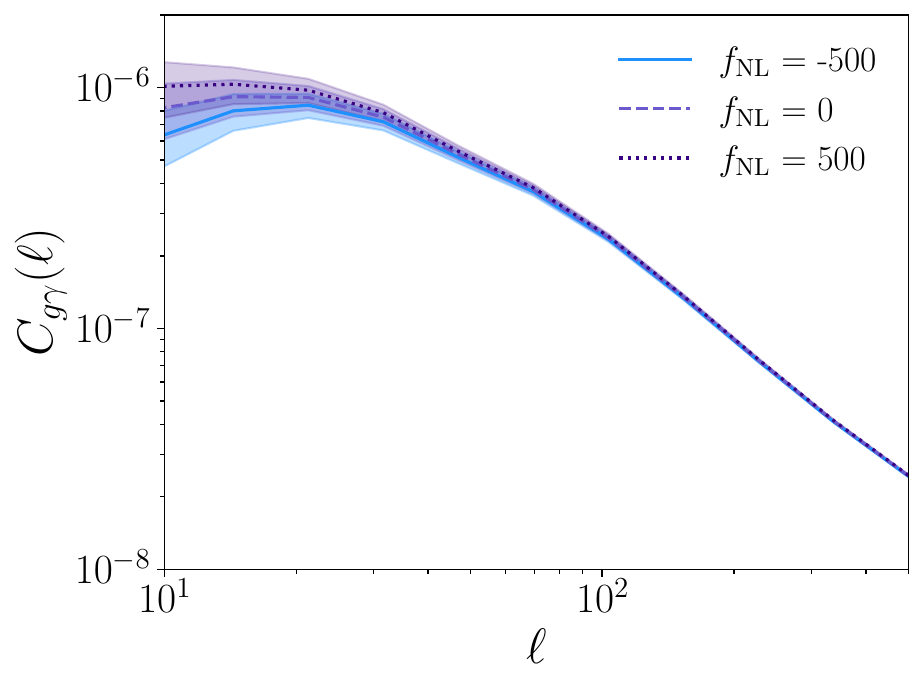}
\includegraphics[width = 0.32\textwidth]{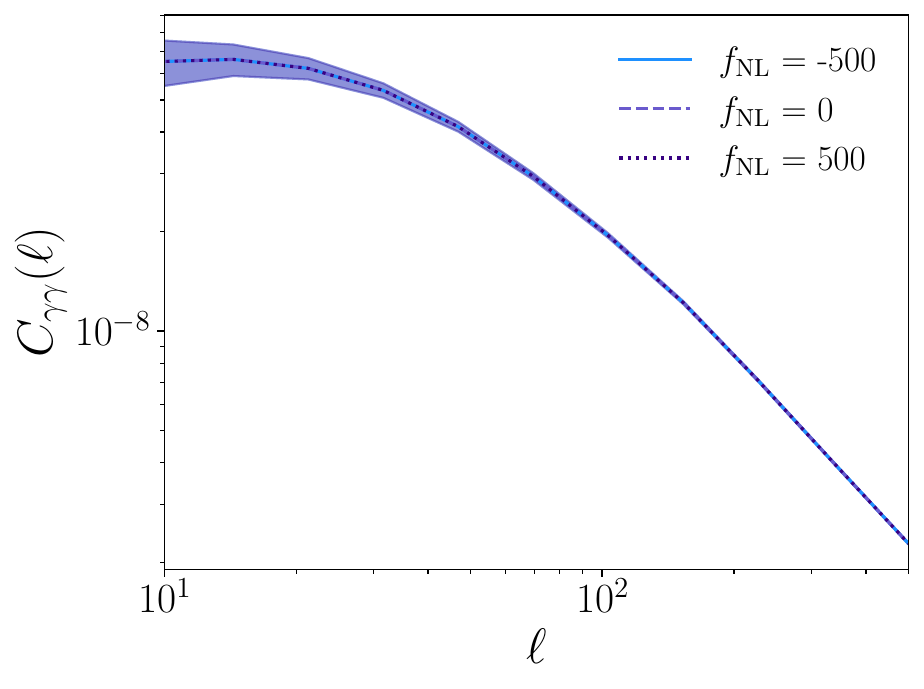} \hfill
\end{center}
\caption{\label{fig:fnl_ex} Lens-galaxy angular correlation function (left), galaxy-shear cross-correlation (middle), and shear-shear correlation (right), for $\Lambda$CDM models with 3 different values of $\fnl$, for galaxies in lens bin 3 and source bin 5 of the LSST Y10 3x2pt setup (see Section~\ref{sec:LSST setup}). 
Extreme values of $\fnl$ are used here to illustrate the nature of the change in $C(\ell)$ shape; error bands are from Eqn. \ref{eq:cov}.  We use a scale range of $10<\ell<500$ divided into 10 log-spaced bins.
The correlations are sensitive to $\fnl$ through the latter's effect on the large-scale galaxy bias. As a consequence, the shear two-point correlation function has no sensitivity to this parameter as it does not use the galaxy field; see text in Section \ref{sec:fnl_into} for a detailed discussion.}
\end{figure*}

Our analysis of PNG models uses a modified version of the \texttt{jax\_cosmo} library that includes the $\fnl$ signatures on the bias above\footnote{\url{https://github.com/DhayaaAnbajagane/jax_cosmo.git}}. An illustration of this effect on weak lensing and galaxy clustering observables is shown in Fig.~\ref{fig:fnl_ex}, where we see that the effect is most prominent in the galaxy auto-correlation on large scales. The tightest current bounds on $\fnl$ come from \textit{Planck}, with $\fnl$ = $-0.9 \pm 5.1 (1\sigma)$, and current LSS measurements provide $\sigma(\fnl) \sim$25-50. Forecasts indicate that future galaxy surveys have the statistical power to decrease this by 1-2 orders of magnitude \citep{fnl_constraints, InflationTheoryObs}. We extend these past efforts by including other tracer fields, namely the GW sources, and by including weak lensing to self-calibrate the galaxy bias parameters and further constrain the $w$CDM and $\Lambda$CDM cosmological parameters, which in turn improve the marginalized constraints on $\fnl$.

\section{Analysis Setup} \label{sec:setup}

In this section, we specify the different sample characteristics and parameter priors that we use for the forecasts. In Sec.~\ref{sec:LSST setup}, we describe the samples from LSST, in Sec.~\ref{sec:gw setup} we do the same for the samples from GW observations, and in Sec.~\ref{subsec:fisher} we describe the procedure we use to compute angular correlation functions and Fisher matrices and predict the constraining power of each of the setups.

\subsection{Photometric Galaxy Surveys: \textit{The LSST Y10 setup}} \label{sec:LSST setup}

\begin{figure*}
\begin{center}
\includegraphics[width = \textwidth]{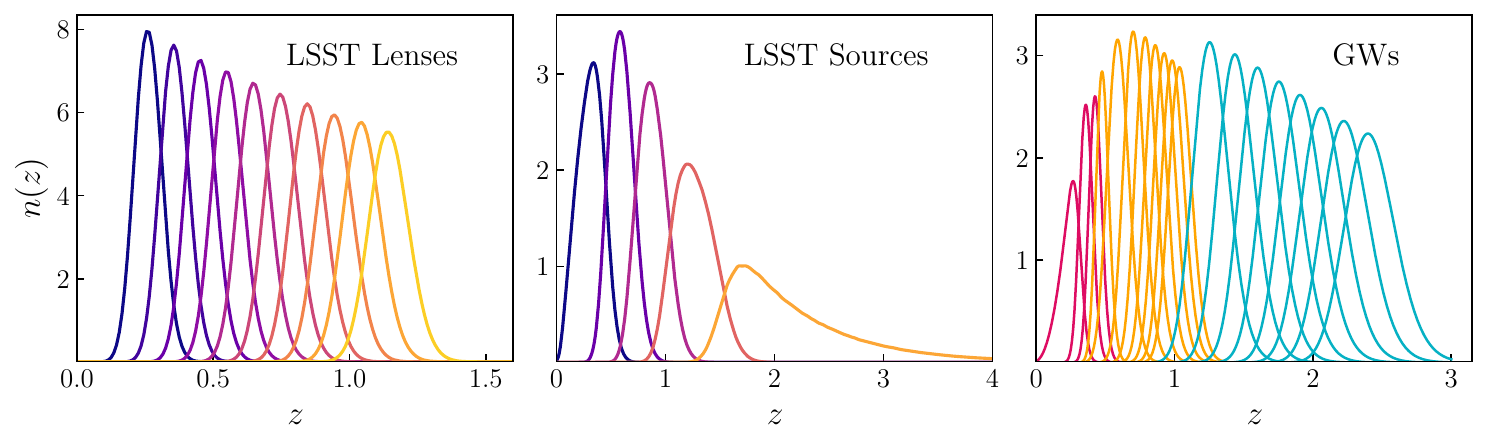}
\end{center}
\caption{$n(z)$ distributions for LSST Y10 lenses, LSST Y10 sources, and the three GW detector network setups. The binning for each GW setup includes the bins of the setup before it (e.g., Setup 3 is composed of the Setup 1 bins (\textcolor{purple}{pink}), the bins added in Setup 2 (\textcolor{orange}{orange}), and the bins added in Setup 3 (\textcolor{teal}{blue}).  The distributions are normalized such that they integrate to unity
and therefore are not representative of the orders of magnitude difference in number density between the LSST and GW samples.}
\label{fig:n(z)s}
\end{figure*}

To generate the 3$\times$2pt data vector we consider a survey corresponding to the LSST Year 10 setup covering 14,300 deg$^2$ of extragalactic (high Galactic latitude) sky, which corresponds to a fraction of the sky $f_{\mathrm{sky}}=0.3466$.  Galaxies are divided into source and lens populations -- sources are objects for which we use both position and shape information, and lenses are objects for which we use only the position information. There is overlap between the two sets of objects, as some galaxies fall into both samples. We show the redshift distributions for each of the 10 lens and 5 source tomographic photo-z bins in the first two panels of Fig.~\ref{fig:n(z)s}. The overall $n(z)$ distribution is modeled as in \citet{Zhang2022}:
\begin{equation}
    \frac{dN}{dz} \propto z^2\rm{exp}[(-z/z_0)^\alpha]~,
\end{equation}
with $(z_0,\alpha)=(0.26, 0.90)$ for lenses and $(0.11, 0.68)$ for sources.  The distributions are separated into 10 equipopulated lens bins and 5 equipopulated source bins, which are convolved with Gaussians of width  corresponding to the expected photo-$z$ precision -- $0.03(1+z)$ for lenses and $0.05(1+z)$ for sources. 

In Table~\ref{tab:LSST_bins}, we list the peak redshift, total galaxy surface number density, and  the fiducial value of the linear galaxy bias for each lens redshift bin and the assumed shape noise (intrinsic variation in the ellipticity of source galaxies) for source galaxies. The source bins have been chosen to have the same surface number density in each bin. 
These values are taken from the LSST DESC Science Requirement Document (DESC SRD, \cite{DESCSRD}). We implicitly assume uninformative priors on the cosmological parameters and on the galaxy bias parameters.

In Fig.~\ref{fig:fnl_ex} we show the forecast $\Lambda$CDM 2D angular power spectrum for LSST Year 10 lens bin 3 and source bin 5 for each of the probes we use for the 3$\times$2pt analysis: galaxy clustering ($C_{\rm gg}$), galaxy-galaxy lensing  ($C_{\rm g \gamma}$) and cosmic shear ($C_{\gamma \gamma}$).

For the uncertainties, we assume a Gaussian covariance matrix, the diagonal elements ($i=j$) of which are given by
\begin{subequations}
\begin{equation}
    \mathsf{Cov}(C_{\rm gg}(\ell)) = \frac{2}{(2\ell + 1)f_{\mathrm{sky}}\, \Delta \ell}\bigg{(} C_{\rm gg}(\ell) + \frac{1}{n_{\mathrm{gal}}} \bigg{)}^2,
\end{equation}
\begin{equation}
    \mathsf{Cov}(C_{\rm \gamma\gamma}(\ell)) = \frac{2}{(2\ell+1)f_{\mathrm{sky}} \, \Delta \ell}\bigg{(} C_{\rm \gamma\gamma}(\ell) + \frac{\sigma_e^2}{n_{\mathrm{gal}}}\bigg{)}^2,
\end{equation}
\begin{equation}
\begin{split}
    \mathsf{Cov}(C_{\rm g\gamma}(\ell)) = \frac{1}{(2\ell+1)f_{\mathrm{sky}}\, \Delta \ell}\bigg{[} C_{\rm g\gamma}(\ell)^2
    \\
     + \bigg{(}C_{\rm gg}(\ell) + \frac{1}{n_{\mathrm{gal}}}\bigg{)}\bigg{(}C_{\rm \gamma\gamma}(\ell)+\frac{\sigma_e^2}{n_{\mathrm{gal}}}\bigg{)}\bigg{]},
\end{split}
\end{equation}
\label{eq:cov}
\end{subequations}
where $\Delta\ell$ is the width of the chosen $\ell-$binning, $n_{\rm gal}$ is the surface number density of lens or source galaxies in the $i$th redshift bin, and $\sigma_e$ is the shape noise for the source galaxies (see Eq. 16-23 in \cite{Friedrich_2021}). We generate angular power spectra and compute the covariance matrix using the $\jax$ library.

\begin{figure*}
\begin{center}
\includegraphics[width = 0.48\textwidth]{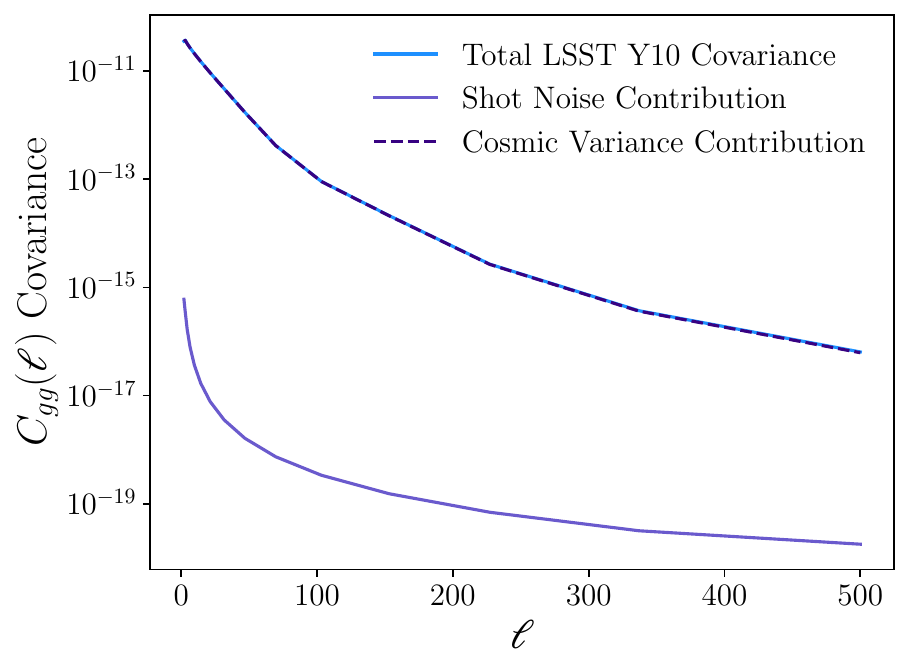}
\qquad
\includegraphics[width = 0.48\textwidth]{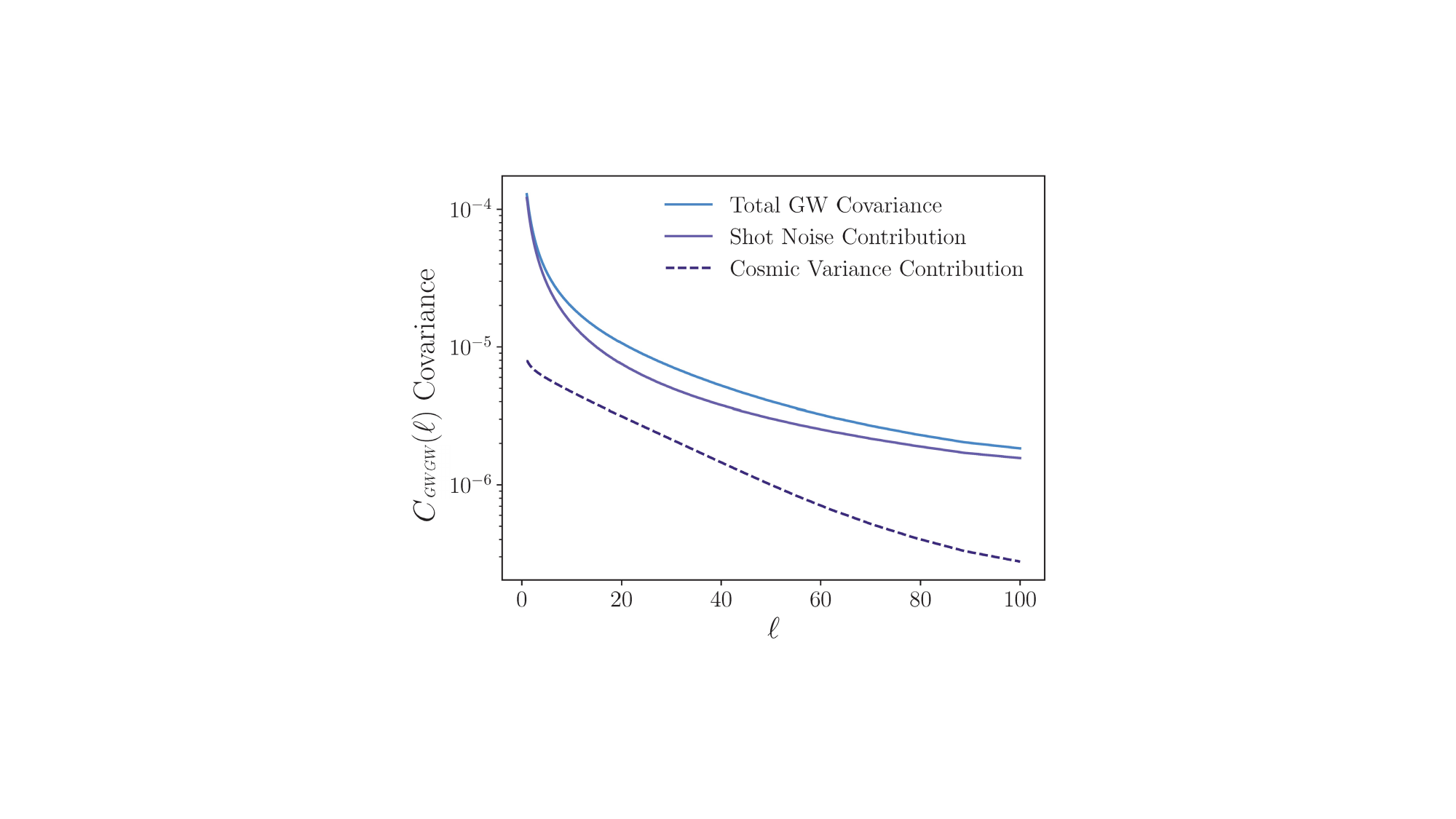} \hfill
\end{center}
\caption{Galaxy clustering covariance for LSST lens bin 7 (\textit{left}) and for the optimistic (Setup 3) GW bin 6 (\textit{right}).  Cosmic variance is the dominant source of covariance for the LSST Y10 lens bin due to the high number density in the survey. The number density of GW events is orders of magnitude lower, so that shot noise dominates and leads to an overall  greater covariance. \label{fig:covs}}

\end{figure*}

The covariance of the angular power spectra comes from two effects: cosmic variance and shot or shape noise. The cosmic variance term is proportional to the signal, $C(\ell)$, and increases at larger scales. Shot noise in galaxy clustering and shape noise in cosmic shear are inversely proportional to galaxy number density $n_{\rm gal}$ -- as the number density of observed objects increases, the shot/shape noise component decreases. In the left panel of Fig.~\ref{fig:covs} we show the diagonal elements of the gg covariance matrix for LSST Y10 lens bin 7, with the separate shot noise and cosmic variance terms. For LSST Y10 lenses, cosmic variance dominates by several orders of magnitude over shot noise for all scales of interest, due to the relatively high number density (large depth) of the sample.

\begin{table}
    \begin{center}
    \textbf{LSST Y10 Baseline Setup}\\
    \begin{tabular}{c c c c c c}
    \hline
    Bin & $z_{\rm peak}$ & Number Density & Galaxy Bias & $\ell_{\rm min}$ & $\ell_{\rm max}$ \\
    & & [arcmin$^{-2}$] &  &  &  \\
    \hline
    Lens &  & &  & &  \\
    1 & 0.255 & 2.63 & 1.09 & 50 & $217$\\
    2 & 0.355 & 3.54 & 1.15 & \same &$294$\\
    3 & 0.455 & 4.1  & 1.21 & \same &$369$\\
    4 & 0.545 & 4.32 & 1.27 & \same &$432$\\
    5 & 0.645 & 4.29 & 1.33 & \same &$499$\\
    6 & 0.745 & 4.08 & 1.4  & \same &$562$\\
    7 & 0.845 & 3.76 & 1.46 & \same &$622$\\
    8 & 0.945 & 3.39 & 1.53 & \same &$679$\\
    9 & 1.045 & 2.99 & 1.6  & \same &$732$\\
    10 & 1.145 & 2.6 & 1.67 & \same &$783$\\
    \hline
    Source &   & Shape Noise &  & \\
    1 & 0.335 & 5.4 & 0.26 & 50 &500\\
    2 & 0.585 & \same & \same & \same & \same\\
    3 & 0.855 & \same & \same & \same & \same\\
    4 & 1.195 & \same & \same & \same & \same\\
    5 & 1.675 & \same & \same & \same & \same\\
    \hline
    \end{tabular}
    \caption{LSST Y10 baseline setup. $\ell_{\rm max}$ values for the lenses derive from the small-scale analysis cut $k_\mathrm{max} = 0.3h$/Mpc.  
    Number density and shape noise are constant across source-galaxy bins by design.
    \label{tab:LSST_bins}}
    \end{center}
\end{table}

\subsubsection{Range of scales} \label{sec:scales_lsst}

\begin{table}
    \begin{center}
    \begin{tabular}{c c c }
    \hline
    & Large-scale cuts & Effective $\ell_{\rm min}$ \\
    \hline
    DES Y3 \citep{desy3-3x2} &&\\
     $\xi_{+}(\theta)$ & $\theta<250$' & 43 \\
     $\xi_{-}(\theta)$ & $\theta<250$' & 43 \\
     $\gamma_{t}(\theta)$ & $\theta<250$' & 43 \\
     $w(\theta)$ & $\theta<250$' & 43 \\
     \hline
   KiDS 1000 \citep{kids1000-3x2} & &  \\
    $C_{\gamma \gamma} (\ell) $  & $\ell > 100 $ & 100  \\
    $C_{\gamma g} (\ell) $  & $\ell > 100 $ & 100 \\
    \hline
   HSC Y3 \citep{hsc-3x2} & &  \\
   $\xi_{+}(\theta)$   & $\theta<50$'  & 216\\
    $\xi_{-}(\theta)$  & $\theta<105$' & 103 \\
    $\Delta \Sigma (R)$  & $R<30-80 h^{-1}$ Mpc & 45  \\
    \hline
    \end{tabular}
    \caption{Summary of large-scale cuts used in recent Stage III 3$\times$2pt analyses and the corresponding values of $\ell_{\rm min}$. The range in $\ell_{\rm min}$ values in the bottom entries for KiDS and HSC come from converting fixed length scale cutoff $R$ to corresponding angular multipole for different redshift bins.   
    \label{tab:scales_summary}}
    \end{center}
\end{table}

As discussed in the Introduction, cosmological analyses of LSS cover a limited range of length or angular scales. Table~\ref{tab:scales_summary} summarizes the large-scale cuts that have been used in recent LSS analyses of Stage III surveys. These cuts are imposed for two main reasons: (i) given the limited sky coverage of these surveys, especially for HSC and KiDS, there are few to no accessible modes on very large scales, leading to very large uncertainties; (ii) observational systematics due to Galactic foregrounds and varying observing conditions have been shown to have the biggest impact on the largest scales. There is currently ongoing effort to develop  methodological improvements to mitigate and control the impacts of these systematic effects on large scales. If this campaign does not succeed, then large scales will need to be discarded or downweighted in upcoming analyses.

Given the Stage III analyses, we choose $\ell_{\rm min}=50$ as the fiducial value for the LSST 3$\times$2pt forecasts when we combine with GW clustering at very large scales. This cutoff corresponds approximately to the maximum of 250 arcmin that was used in the DES Y3 3$\times$2pt analysis. In the analysis below, we also explore both larger and smaller $\ell_{\rm min}$ values. Larger values could potentially be needed if the corrections for systematic effects on large scales are not found to be robust enough, particularly given the smaller statistical errors expected from LSST. We also explore the constraining power for lower $\ell_{\rm min}$ values to understand the gain that would be possible from improved control of systematic effects on large scales.

On small scales, we choose the $\ell_{\rm max}$ values (see Table~\ref{tab:LSST_bins}) for the galaxy clustering and galaxy-galaxy lensing probes following the harmonic space scale cuts from the DESC SRD \citep{DESCSRD}. This corresponds to a minimum physical scale cut of $k_\mathrm{max} = 0.3h$/Mpc, limited by the modeling of non-linearities and baryonic effects in the galaxy power spectrum. For cosmic shear, the DESC SRD used $\ell_{\rm max}=3000$, which is more aggressive than previous work; we choose to adopt a more conservative cut of $\ell_{\rm max}=500$ for our fiducial analysis. Later we explore the impact of extending the LSST analysis to smaller length scales.

\subsection{Gravitational Wave Samples} \label{sec:gw setup}

For the GW samples, we include gravitational waves produced by coalescing binary black holes, black hole/neutron star binaries, and binary neutron stars. We expect these events to occur in galaxies, as they are home to the most intense star formation, implying that the distribution of GWs is expected to follow that of the underlying dark matter large-scale structure \citep{Bosi2023}. Since the LSST analysis setup is already well-defined --- see Section \ref{sec:LSST setup} above, and also \citet{DESCSRD} --- our goal here is to determine the type of GW experiment/sample that best complements the expected measurements from LSST. As a result, we intentionally consider simplistic setups for the GW experiments, with some setups approaching highly futuristic assumptions. Our goal here is to determine the impact of supplementing information from galaxy surveys with that from GW experiments, and so our results are not precise predictions of constraints for a given, \textit{planned/ongoing} GW experiment.

Following the approach of simplicity, motivated by the above goal, we describe all GW samples using only three parameters, all of which are determined by the chosen detector setups: the number of GW detections per year, their sky localization precision, and the redshift range of the detected sources. The number of detections $n_{\rm GW}$ determines the shot noise contribution to the clustering power spectrum -- the number density of predicted GW events is orders of magnitude lower than that of galaxies in photometric surveys, which leads to a higher contribution from shot noise. The precision with which GW events can be localized on the sky determines the minimum angular scale (maximum $\ell$) on which the clustering of GW events can be inferred. Lastly, the redshift range (and precision) determines the cosmological information contained in the GW source clustering. Redshift range overlap between the GW and LSST sources is beneficial, as it potentially strengthens the constraints on galaxy bias for both sets of sources. In addition, for future GW detector networks with high sensitivity, a large fraction of GW events are predicted to be concentrated at higher redshifts -- the peak of the forecast GW $n(z)$ distribution is at $z \sim 2$, since this is the redshift at which the star formation rate peaks, and a significant fraction of events occurs at even higher redshift.

We define three GW setups of increasing optimism, with different values for the detector network parameters $n_{\rm GW}$, $\ell_{\rm max}$, and $z$ range. The values of these parameters are listed in Table~\ref{tab:GW_bins}, and the corresponding redshift bins are shown in the right panel of Fig.~\ref{fig:n(z)s}.

Setup 1 is the least optimistic, as it assumes a relatively low number density of events, moderate spatial localization ($\ell_{\rm max}$) and limited $z$ range. The values of these parameters were chosen to approximately correspond to a third generation (3G) GW detector network comprising the Einstein Telescope \citep[ET,][]{ET_science_case}, Cosmic Explorer \citep{CE_reference}, and LIGO Voyager \citep{LIGOVoyager}.
Such a network is expected to have a conservative localization precision such that  $\ell_{\rm max}=30$ over the redshift range $0<z<0.5$ (see Fig. 4 of \citealt{hall&evans}), a median luminosity distance uncertainty of $\sim13$\%, and $10^4$ detections/year \citep{Borhanian2022}. We also check below (in Figure \ref{fig:FoM fnl results}) the sensitivity of our results to the choice of $\ell_{\rm max}=30$ in this setup, and will show that changing this by a factor of three does not change the combined LSST plus GW analysis.

Setup 3 considers the ideal case in which all events are detected with extremely good localization; it is approximately based on a network composed of three ET-like third generation telescopes. We use a GW detection rate of $10^6$ events/year, with a $\sim5$\% uncertainty in luminosity distance (\cite{ET_science_case}, Fig. 12). Following the specifications in \cite{Namikawa2016}, we define Setup 3 to cover the redshift range $0<z<3$, with $\ell_{\rm max}=90$ over the redshift range $0<z<1$ and $\ell_{\rm max} = 57$ for $1<z<3$. We then consider an additional case, Setup 2, as a middle scenario between 1 and 3, with $10^5$ detections/year, redshift range of $0<z<1.3$, and localization allowing for analysis of scales with $\ell<60$ over this range. This setup aligns well with current expectations for next-generation detectors. As shown in Figure 3 of the recent review by \citet{Chen_2024}, the expected cumulative number of BBH detections for Cosmic Explorer (CE) and Einstein Telescope (ET) reaches about $10^5$ per year by the late 2030s.

The redshift binning of Setup 1 comprises three bins over the range $0<z<0.5$.  Setup 2 contains the three bins of Setup 1, as well as an additional nine bins covering the range $0.4<z<1.3$.  Setup 3 is composed of the twelve Setup 2 bins and eight more bins over the range $1<z<3$. In all cases, our choice of binning is informed by the uncertainty in the luminosity distance for a given setup. We convert luminosity distance to redshift assuming a fiducial cosmology (values listed in Section 2.1).
The right panel of Fig.~\ref{fig:n(z)s} shows these distributions (the details of how they are constructed are given below). Although the smaller luminosity distance uncertainty we assume in Setup 3 would allow us to use finer redshift binning over the range $1<z<3$, testing has shown that in our analysis a greater proportion of cosmological information is contained in the $z\lesssim 1$ range rather than in the higher redshift ranges. Additionally, overlap between the GW and LSST redshift bins improves constraining power, and the LSST redshift distributions reside mostly in the $0<z<1.5$ range.  A larger number of GW redshift bins also increases computational complexity and decreases the number density of objects in each bin; the three sets of GW $n(z)$ bins we propose represent a optimal middle-ground between fine binning of the lower-$z$ range and computational complexity.

To generate the total GW redshift distribution, we assume that the number density of GW sources is proportional to the star formation rate described in \cite{Madau&Dickinson}. The star formation rate function $R_{\star}(z)$ is used to compute the number of events expected at varying masses and redshifts:
\begin{equation}\label{eqn:Rz_SFR}
    R_{\star}(z) = \frac{(1+z)^{2.6}}{1+\left(\frac{1+z}{3.2}\right)^{6.2}}, \qquad
    \frac{dn}{dz} = A_\star R_{\star}(z) \, \frac{V_c(z)}{1+z},
\end{equation}
where $V_c$ is the differential comoving volume and $A_\star$ normalizes to the fraction of stellar mass that is converted to GW sources and our chosen number of detections per year. In practice, $A_\star$ is set by the number of detections we assume for a given experiment (see Section \ref{sec:gw setup} for details on the different experiment setups). It is relevant to note that the GW formation channel affects the time delay between progenitor star formation and binary coalescence, and therefore the shape of the GW $n(z)$ distribution.  Our model assumes this time delay is relatively short (< 1Gyr), which \citet{Fishbach_2021} find is favored in all formation channels they consider. The form of Equation \eqref{eqn:Rz_SFR} is further motivated in \citet{Vijaykumar_2023}, who constrain different GW host galaxy trends using data from the GW transient catalog 3 (GWTC-3) and find that a majority of GW events originate in galaxies of high SFR; see their Figure 1 and Section 2. This motivates using the cosmic star-formation rate to define the shape of the $n(z)$, as we have done here.

Additionally, we assume that in the $z$ ranges of our setups, the detection rate of events is not dependent on redshift. GW events within these ranges have a signal-to-noise ratio above the detection threshold for all networks considered \citep{Borhanian2022}, therefore we assume a fixed fraction is detectable independent of redshift.

The binning choice of the GW redshift distribution is dependent on the parameters of the chosen detector setup. We divide the redshift range for a particular setup into $n_{\rm bins}$ subintervals such that each subinterval contains an equal number of GW events drawn from the total $n(z)$ distribution. To transform these true redshift distributions into the observed distributions, we convolve the distribution in each subinterval with a Gaussian of width $\sigma(z)$ determined by the detector's redshift uncertainty as inferred from the above uncertainty in luminosity distance. This generates the set of $n_{\rm bins}$ redshift distributions shown in Fig.~\ref{fig:n(z)s}.

Number density is determined based on the detection rate of the chosen network. We calculate the fraction of total events contained in each $z$ range and divide by the number of bins within that range, giving the fraction of events contained within each individual (equipopulated) bin. Multiplying by the yearly detection rate with ten years of observation and dividing by the observed area (in this case, the full sky) gives us the number density of events per bin. For use with $\jax$'s differentiable functionality, we convert the binned distributions to smail approximations described by smooth functions. The area under the curve of each bin is normalized to unity and then multiplied by the number density scalar.

We also assume that all experiments can access the full-sky, i.e. they have $f_{\rm sky} = 1$. As discussed before, GW experiments are not limited to observing only certain regions of the sky, as is the case with optical surveys where the region of sky adjacent to the galactic plane is unused/unobserved. However, GW experiments are more sensitive to certain regions of the sky than others, and so in reality, the effective observed sky-fraction will be $f_{\rm sky} < 1$; the specific value depends on the exact positions of the experiments in the network. Following our goal to study simplistic/best-case setups, we will use $f_{\rm sky} = 1$. Lowering the observed sky fraction can also affect the largest scales used in the analysis --- which is pertinent as one of our signals, the scale-dependent galaxy bias, peaks on larger scales --- so we will also show results from varying this choice for the largest usable scale (Table \ref{tab:verification_table} and Section \ref{subsubsec:ellmin_GW}).

The fiducial large-scale bias for the GW sources is computed for the median redshift of each bin using this parameterization
\begin{equation}
    b_{\rm GW} = 1.20(1+z)^{0.59} ~,
    \label{eq:GW_bias}
\end{equation}
from \citet{Peron2023}, who extract tracer power spectra from mock data-sets of GW events and fit them to a model of the true power spectrum. The result is the above estimated bias. The power spectrum model used in this case neglects third order and scale-dependent contributions, assuming only contributions from the local matter field. The same bias parameterization is assumed in all of our setups. Bias evolution with redshift (as well as time delay between star formation and binary coalescence) depends on the formation channel that leads to GW events, which is currently uncertain -- in Section \ref{subsubsec:GW_bias} we test the impact of other choices for the GW bias. Note that each redshift bin of the GW sample  is modelled using a separate GW bias parameter; we \textit{do not} use a fixed bias relation for our analysis. All cosmology results below are obtained after varying, and marginalizing over, these bias parameters. Therefore, the relation in Equation \eqref{eq:GW_bias} only defines the fiducial bias values around which we compute derivatives for estimating the Fisher matrix.

\begin{table}
    \begin{center}
    \textbf{GW Setup}\\
    \begin{tabular}{c c c c c}
    \hline 
    Bins & Number Density & $\ell$ range & $z_{\rm peak}$ & $b_{\rm{GW}}$\\
    & [arcmin$^{-2}$] &&&\\
    \hline
    Setup 1 &&&&\\ 
    \textcolor{purple}{1} & $1.14\times10^{-5}$ & $2<\ell<30$ & 0.27 & 1.31\\
    \textcolor{purple}{2} & \same & \same & 0.36 & 1.44\\
    \textcolor{purple}{3} & \same & \same & 0.43 & 1.48\\
    \hline
    Setup 2 & Number Density & $\ell$ range & $z_{\rm peak}$ & $b_{\rm{GW}}$\\
    \textcolor{purple}{1-3} & $1.14\times10^{-4}$ & $2<\ell<60$ & & \\
    \textcolor{orange}{4} & $1.6\times10^{-4}$ & $2<\ell<60$ & 0.48 & 1.57\\
    \textcolor{orange}{5}  & \same & \same & 0.59 & 1.60\\
    \textcolor{orange}{6}  & \same & \same & 0.70 & 1.64\\
    \textcolor{orange}{7}  & \same & \same & 0.79 & 1.69\\
    \textcolor{orange}{8}  & \same & \same & 0.86 & 1.73\\
    \textcolor{orange}{9}  & \same & \same & 0.93 & 1.77\\
    \textcolor{orange}{10} & \same & \same & 0.98 & 1.80\\
    \textcolor{orange}{11} & \same & \same & 1.04 & 1.83\\
    \textcolor{orange}{12} & \same & \same & 1.25 & 1.94\\
    \hline
    Setup 3 & Number Density & $\ell$ range & $z_{\rm peak}$ & $b_{\rm{GW}}$\\
    \textcolor{purple}{1-3} & $1.14\times10^{-3}$ & $2<\ell<90$ & \\
    \textcolor{orange}{4-12} & $1.6\times10^{-3}$ & $2<\ell<90$ & \\
    \textcolor{teal}{13} & $4.2\times10^{-3}$ & $2<\ell<57$ & 1.43 & 2.03\\
    \textcolor{teal}{14}  & \same & \same & 1.60 & 2.11\\
    \textcolor{teal}{15}  & \same & \same & 1.75 & 2.19\\
    \textcolor{teal}{16}  & \same & \same & 1.91 & 2.26\\
    \textcolor{teal}{17}  & \same & \same & 2.06 & 2.33\\
    \textcolor{teal}{18}  & \same & \same & 2.22 & 2.50\\
    \textcolor{teal}{19}  & \same & \same & 2.40 & 2.58\\
    \hline
    \end{tabular}
    \caption{Binning, GW source number density, $\ell-$range, peak redshift, and fiducial GW host bias for the three GW setups. 
    In Setup 3, the $\ell-$range for lower-$z$ bins is extended to $2<\ell<90$, whereas the higher-$z$ bins only go up to $\ell$ of 57.  Number densities are normalized to ten years spent observing.  $n(z)$ distributions are plotted in Fig.~\ref{fig:n(z)s}.
    \label{tab:GW_bins}
    }
    \end{center}
\end{table}

We generate the GW clustering angular power spectrum using the same \texttt{jax-cosmo} framework as for LSST. This same software also computes the cross-covariance between the GW $C(\ell)$ arrays and the $3\times$2pt probe from LSST.

\subsection{Forecasting procedure} \label{subsec:fisher}

We perform a Fisher forecast for the constraining power of the different probes under different setups. The Fisher matrix elements in parameter space are defined by
\begin{equation}
    \textbf{F}_{mn} = \sum_{l_1,l_2} \frac{dC(\ell)_{l_1}}{dp_m}(\mathbf{C}^{-1})_{l_1l_2}\frac{dC(\ell)_{l_2}}{dp_n},
\end{equation}
where $(\mathbf{C}^{-1})$ is the inverse of the covariance matrix, and $dC(\ell)_{l_1}/dp_m$ is the derivative at point $l_1$ in the $C(\ell)$ data vector (which includes all angular and cross power spectra, including across redshift) with respect to parameter $p_m$. $\textbf{C}$ is a simple Gaussian covariance matrix, as defined in Section \ref{sec:LSST setup}, and is computed with the angular power spectra in the $\jax$ codebase, and $dC(\ell)/dp_m$ is computed using the auto-differentiation capabilities of $\jax$ functions. The covariance matrix for a GWxLSST setup is shown in Fig. \ref{fig:cov_appendix} of Appendix A. In the fiducial case, \textbf{F} has dimension $(n_{\text{cosmo params}} + n_{\text{LSST lens bins}} + n_{\text{GW bins}})^2$, since there is an undetermined galaxy bias parameter for each LSST lens and GW redshift bin, and  $n_{\text{cosmo params}}=5$ ($\Lambda$CDM) or 6 ($w$CDM or $\fnl$ extension). We do not consider other nuisance parameters, such as those related to the $n(z)$ distributions or the shear multiplicative bias. Thus, our results assume those systematics are subdominant.

We construct the Fisher matrix from Jacobian and covariance matrices. Since we use the differentiable \texttt{jax-cosmo} framework to generate the angular power spectra, it is trivial to perform these derivatives in a stable way.  The $(l_1,m)$th term of the Jacobian matrix is defined as
\begin{equation}    
    \mathbf{J}_{l_1m} = \frac{d C(\ell)_{l}}{d p_m},
\end{equation}
for the $l$th value in the $C(\ell)$ array and varied parameter $p_m$.

We first compute one covariance and one Jacobian matrix containing all LSST Y10 lens, LSST Y10 source, and GW probes, spanning the range $2<\ell<3000$ and using an $f_{\rm sky}$ of 1.  We then apply cuts to remove data points/indices corresponding to LSST lens and source bins that do not fall within the actual $\ell$ cuts given in the tables above. The covariance matrix terms of all LSST-associated data points assume $f_{\rm sky}=0.3466$ whereas those of GW-associated data points assume $f_{\rm sky} = 1$.

We compute the Figure of Merit (FoM) as
\begin{equation}
    \text{FoM}_\theta = \cfrac{1}{\sqrt{\det\big{[}(\textbf{F}^{-1})_\theta \big{]}}},
\end{equation}
where $\textbf{F}^{-1}$ is the inverted Fisher matrix of all parameters varied in the analysis (cosmological and galaxy bias), and $\theta$ represents the selection of parameters to be considered in the FoM.


\begin{figure*}
\begin{center}
\includegraphics[width = 0.6\textwidth]{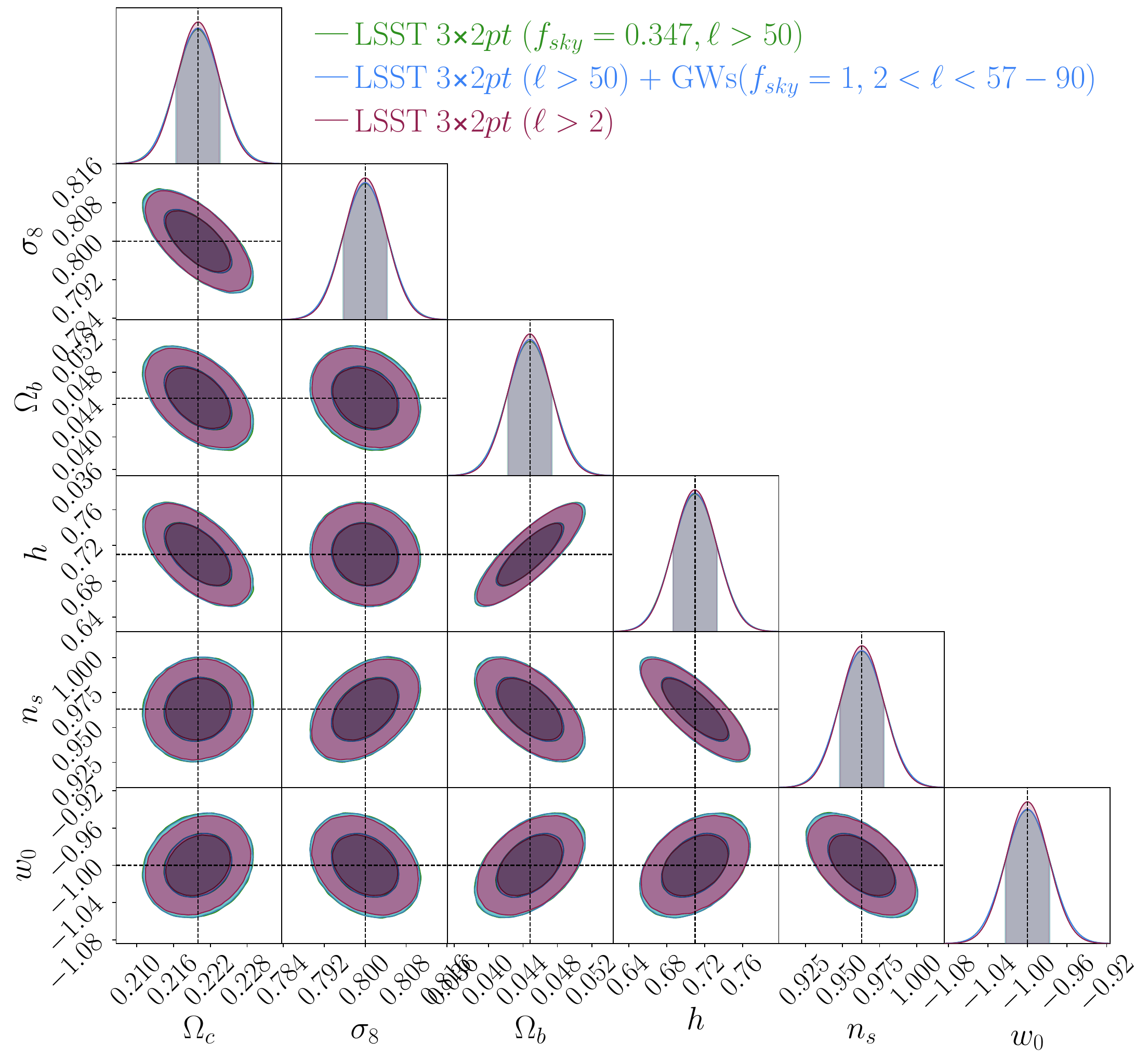}
\end{center}
\caption{Marginalized 68\% and 95\% CL $w$CDM constraints from LSST 3$\times$2pt baseline analysis (green), LSST 3$\times$2pt extended to very large scales (purple), and LSST baseline plus GW auto-correlation on very large scales (light blue). Differences between these are quantified in Table \ref{tab:percent_improvement}.
\label{fig:wcdm_results}}
\end{figure*}

\section{Results} \label{sec:results}

In this section, we present the main results from this study. 
Sec.~\ref{subsec:LCDM_results} discusses $\Lambda$CDM and $w$CDM results for the baseline LSST 3$\times$2pt analysis, for the LSST analysis extended to very  large scales, and for the baseline LSST analysis augmented by GW clustering on very large scales. 
Sec.~\ref{subsec:fNL_results} extends the results to PNG and constraints on $\fnl$. In Sec.~\ref{subsec:analysis_choices}, we vary some of our analysis choices to determine the impact of these setup changes and assumptions on our results. These variations include the LSST scale cuts (\ref{subsubsec:ellmin_ellmax_LSST}) as well as GW galaxy bias (\ref{subsubsec:GW_bias}) and GW scale cuts (\ref{subsubsec:ellmin_GW}).

\subsection{$\Lambda$ and \textit{w}CDM} \label{subsec:LCDM_results}

Here we present the results for the $\Lambda$CDM and $w$CDM models. For now, we only consider the most optimistic case for the GW detector network (GW setup 3).

As an example, Fig. ~\ref{fig:wcdm_results} shows results for 3 different setups in the context of the $w$CDM model: the baseline LSST Y10 3$\times$2pt analysis (green), the LSST analysis extended to very large scales, $\ell>2$ (purple), and the LSST baseline combined with GW source clustering in setup 3 (light blue). The relative gains in precision of cosmological constraints for the latter two cases relative to the LSST baseline are given in Table~\ref{tab:percent_improvement} for both $\Lambda$CDM and $w$CDM. For $w$CDM, the addition of the very large scales ($2<\ell<50$) to the baseline LSST Y10 3$\times$2pt analysis leads to a fractional improvement in the parameter constraints of between 2.8 and 6.2\%. The $\Lambda$CDM results are qualitatively similar. Note that these results assume that residual systematic effects on very large scales ($\ell < 50$) in LSST will be negligible. 

By comparison, combining the baseline LSST analysis with GW clustering at large scales improves parameter constraints over the baseline LSST analysis by less than 1\%. This difference is not surprising, as even in the optimistic GW setup, the shot noise is much larger than for LSST (see Fig. \ref{fig:covs}). However, the neglect of systematic effects in the very large-scale GW measurements appears to be better justified than for LSST.

\begin{table*}
    \begin{center}
    \textbf{Percent Improvement on Parameter Constraints}\\
    \begin{tabular}{c c c c c c c c c c c c c c}
     & \multicolumn{4}{c}{$\Lambda$CDM} & \multicolumn{4}{c}{$w$CDM} & \multicolumn{4}{c}{$w$CDM: $\ell_{\rm max}^{\rm src}=3000$, $k_{\rm max}^{\rm lens} = 0.6$}\\[5pt]
    \cline{1-1}\cline{3-5}\cline{7-9}\cline{11-13}
    Parameter && $\sigma$(param) & + 3$\times$2pt L.S. & + GWs & & $\sigma$(param) & + 3$\times$2pt L.S. & + GWs & & $\sigma$(param) & + 3$\times$2pt L.S. & + GWs\\
    $\Omega_c$ && 0.0036 & 6.1\% & 0.98\% && 0.0037 & 5.1\% & 0.78\% && 0.0023 & 2.2\% & 0.48\%\\
    $\sigma_8$ && 0.0052 & 3.9\% & 0.36\% && 0.0055 & 3.6\% & 0.19\% && 0.0017 & 2.6\% & 0.32\%\\
    $\Omega_b$ && 0.0023 & 2.6\% & 0.61\% && 0.0027 & 4.9\% & 0.78\% && 0.0020 & 2.3\% & 0.52\%\\
    $h$        && 0.022  & 1.1\% & 0.24\% && 0.0025 & 2.8\% & 0.40\% && 0.014  & 0.87\% & 0.25\%\\
    $n_s$      && 0.014  & 1.4\% & 0.09\% && 0.018  & 4.1\% & 0.30\% && 0.0057 & 0.91\% & 0.04\%\\
    $w_0$      && -      & -     & -      && 0.023  & 5.9\% & 0.40\% && 0.014  & 2.3\% & 0.14\%\\
    LSST $b_{\rm gal}$ bin 1 && 0.0093 & 3.9\% & 0.42\% && 0.010 & 5.0\% & 0.60\% && 0.0043 & 3.58\% & 0.41\%\\
    LSST $b_{\rm gal}$ bin 10 && 0.0097 & 4.8\% & 0.58\% && 0.016 & 7.0\% & 0.83\% && 0.0091 & 3.08\% & 0.28\%\\
    \hline
    \end{tabular}
    \caption{Marginalized 68\% CL parameter constraints ($\sigma$(param)) from LSST Y10 3$\times$2pt baseline ($\ell > 50$) and percent improvement in constraints on parameters with addition of LSST measurements on large scales ($\ell=2-50$, denoted +3$\times$2pt L.S.) or with addition of GW clustering analysis on large scales ($\ell=2$ to $57-90$ with $\ell_{\rm max}$ depending on redshift bin) for $\Lambda$CDM and $w$CDM. For LSST $b_{\rm gal}$, we present statistics for redshift bins 1 and 10 to show the range of values produced. For comparison, last three columns show parameter constraints and improvements when the cosmic shear, galaxy-shear, and galaxy clustering measurements are extended to smaller scales in $w$CDM (see text). 
    \label{tab:percent_improvement}}
    \end{center}
\end{table*}

\begin{figure*}
\begin{center}
\includegraphics[width = 0.6\textwidth]{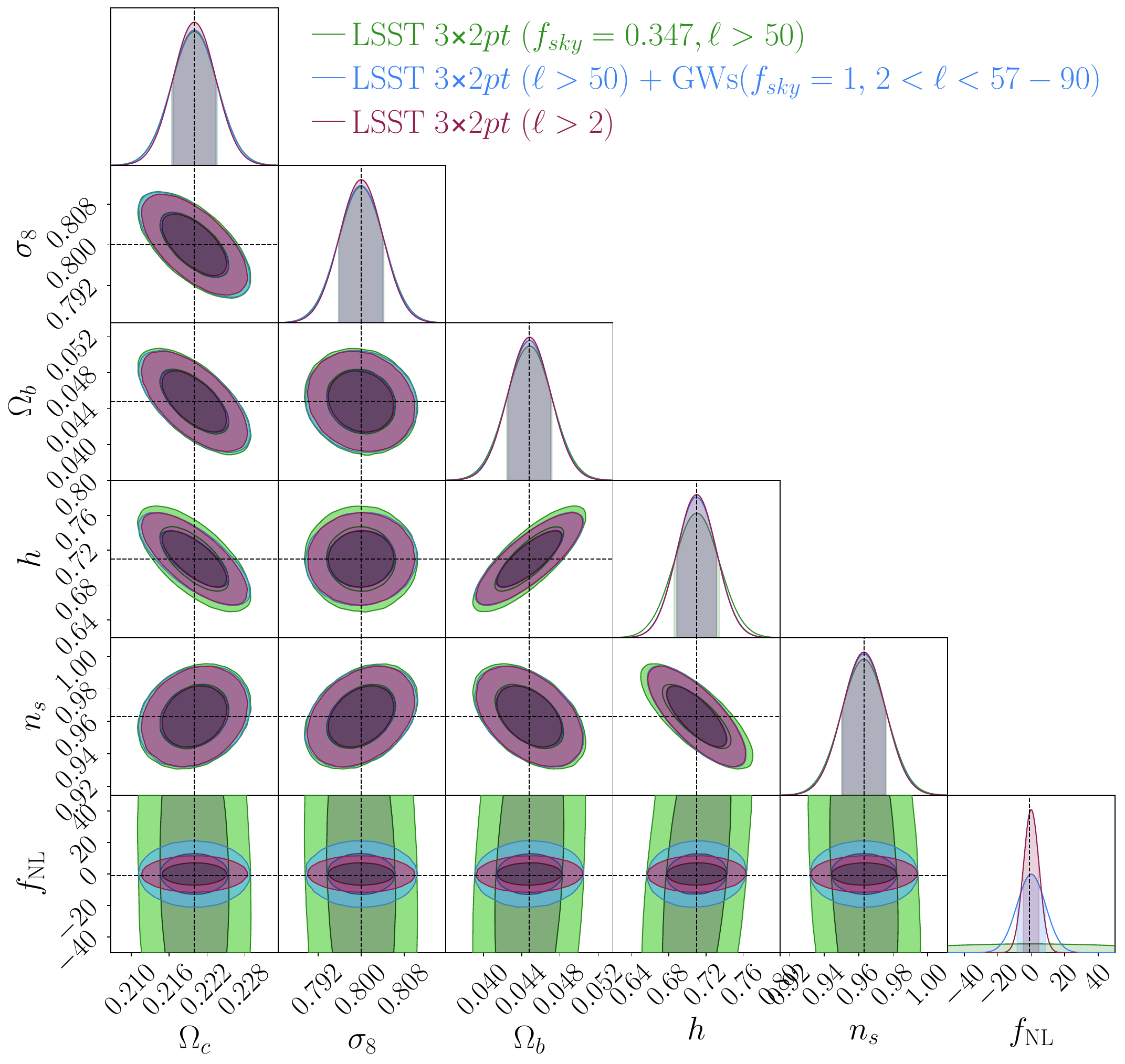}

\end{center}
\caption{Marginalized 68\% and 95\% CL constraints on cosmological parameters in $\Lambda$CDM including PNG parameterized by $\fnl$. The three analysis setups are identical to those in Fig. \ref{fig:wcdm_results}.}  
\label{fig:fnl_contours}
\end{figure*}

Another result of this combined analysis is that in the $\Lambda$CDM model we are able to constrain the 19 GW source galaxy bias values to between 2.7\% (GW bin 1) and 12\% (GW bin 19) precision. This contrasts with a much larger $\sim$218\% precision in GW galaxy bias from GW clustering alone, due to the degeneracy between $\sigma_8$ and GW host bias.  We find similar results for $w$CDM, with precision between 2.8\% (GW bin 1) and 12\% (GW bin 19) from LSST + GW and 245\% from GW alone. This is particularly interesting as constraining the galaxy bias of GW sources is an active area of study and can help us understand the galaxy populations that host GW events and thus GW formation pathways \citep{Vijaykumar_2023, Mukherjee_2021}. 

In the last three columns of Table  \ref{tab:percent_improvement}, we explore the impact on parameter constraints and on the gain from adding very large scale measurements if we adopt more aggressive small-scale analysis cuts in the context of $w$CDM. For cosmic shear, $\ell_{\rm max}^{\rm src}$ is increased from 500 to 3000, and for clustering and galaxy-shear the maximum wavenumber is increased from 0.3 to $0.6 h/$Mpc. The inclusion of smaller scales increases the constraining power of the baseline LSST Y10 analysis, reducing the relative improvement from measurements on very large scales: for $w$CDM, the addition of LSST 3$\times$2pt large scales improves constraints on cosmological parameters by 0.9-2.6\% and addition of GWs by 0.04-0.5\%.  Results for $\Lambda$CDM are similar.

\begin{figure*}
\begin{center}
\includegraphics[width=\textwidth]{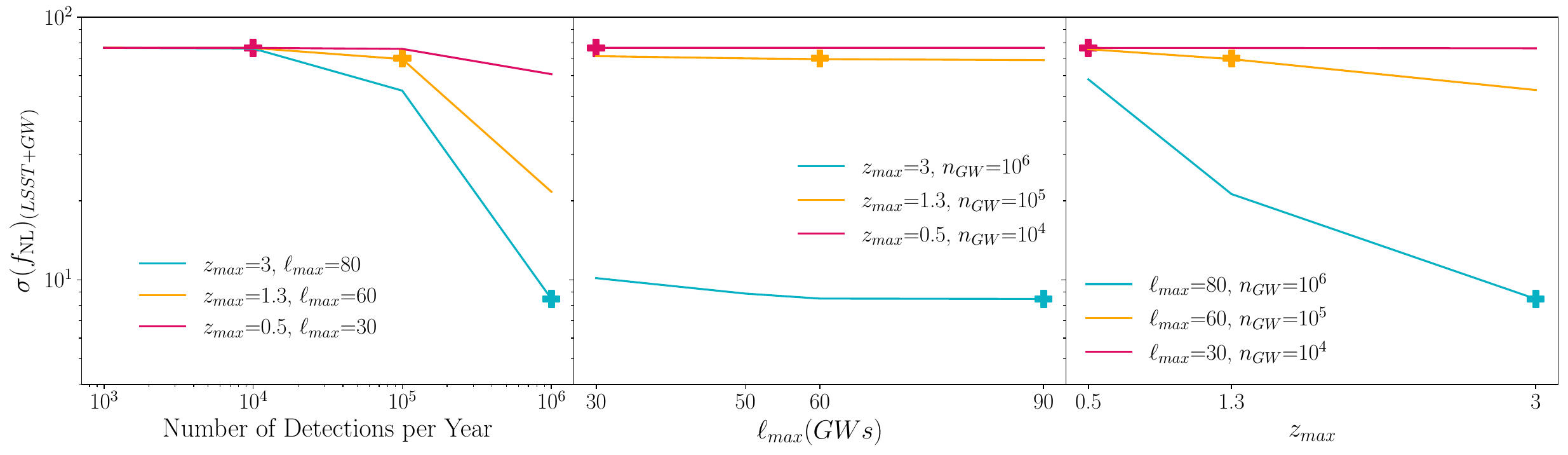}
\end{center}
\caption{Marginalized 68\% CL uncertainty on $\fnl$ with LSST baseline plus varying GW setups.  Points marked with a \ding{58} indicate the base parameters of the 3 GW setups (see text).
}
\label{fig:FoM fnl results}
\end{figure*}

\subsection{Primordial non-Gaussianity and \textit{f}$_{\text{NL}}$} \label{subsec:fNL_results}

Here we present results for the $\Lambda$CDM model with primordial non-Gaussianity characterized by $\fnl$. Using the same setup as in Sec.~\ref{subsec:LCDM_results}, we first consider the most optimistic case for the GW detector network (GW setup 3). We then explore in Sec~\ref{sec:GW_network} how changing the setup affects the results.

\subsubsection{Large scales and $\fnl$} 

Fig.~\ref{fig:fnl_contours} shows the constraints on the $\Lambda$CDM+$\fnl$ parameters from the baseline LSST Y10 3$\times$2pt alone (green), with the addition of GW clustering on large scales (blue), and with the addition of LSST 3$\times$2pt analysis on large scales (purple). The baseline constraints on the $\Lambda$CDM parameters are slightly weaker than those in Table \ref{tab:percent_improvement} due to the marginalization over $\fnl$. 
We see that the addition of information from very large scales ($\ell<50$), either from LSST or GW, dramatically improves the constraint on $\fnl$ -- from $\sigma(\fnl)=76.5$ to $\sigma(\fnl) \simeq 8.5$ for GWs and $\sigma(\fnl)\simeq4.6$ for LSST large scales.
This is not surprising, given that the scale-dependent bias in the PNG model grows at large scale. (Some previous works have found LSST galaxy clustering alone to provide tighter constraints on $\fnl$ \citep[e.g.][]{Dizgah2018LSSTfNL, Green2023LightFields} due to either using a lower $\ell_{\rm min}$ or marginalizing over fewer parameters than we do here.)
Fig.~\ref{fig:fnl_contours} also shows that the addition of very large scale information noticeably improves the constraints on the Hubble parameter $h$; this is discussed further in Sec.~\ref{subsubsec:ellmin_ellmax_LSST}. 

Although the extension of the LSST 3$\times$2pt analysis to very large scales yields tighter statistical constraints than the addition of GW clustering, as noted above we expect the latter to be much more robust to systematic errors.

\subsubsection{Dependence on GW detector network parameters} 
\label{sec:GW_network}

Fig.~\ref{fig:FoM fnl results} shows how the uncertainty on $\fnl$ depends on GW setup. Of the three GW detector network parameters we vary, the number of detections per year and the redshift range significantly impact the constraining power on $\fnl$, while $\ell_{\rm max}(\rm GWs)$ is relatively unimportant. Once $n_{\rm GW}$ increases above $\sim$$10^5$ detections/year, the drop in shot noise is such that the large-scale GW clustering measurement provides a useful constraint in combination with LSST. Additionally, we find that $z_{\rm max}$ of at least $\sim1.3$ is required for GW clustering to add significant constraining power. For reference, the LSST Y10 lenses have $z_{\rm max}$ of $\sim$1.4. This is important for multiple reasons -- having more GW redshift bins allows for more cross-bin correlation and therefore tighter constraints, and higher $z_{\rm max}$ also means more overlap with LSST and a larger fraction of total GW events captured.  At $z_{\rm max}$ of 1.3, GWs cover nearly the entire LSST lens range and most of the LSST source range. However, increasing $z_{\rm max}$ to 3 allows us to include 8 more redshift bins and many more GW events -- the peak of the GW $n(z)$ distribution is around $z=2$. By contrast, we find that better localization of events and therefore higher $\ell_{\rm max}$ has minimal effect on constraining power. This is because most information on $\fnl$ is found at extremely large scales.

\begin{figure*}
\begin{center}
\includegraphics[width = \textwidth]{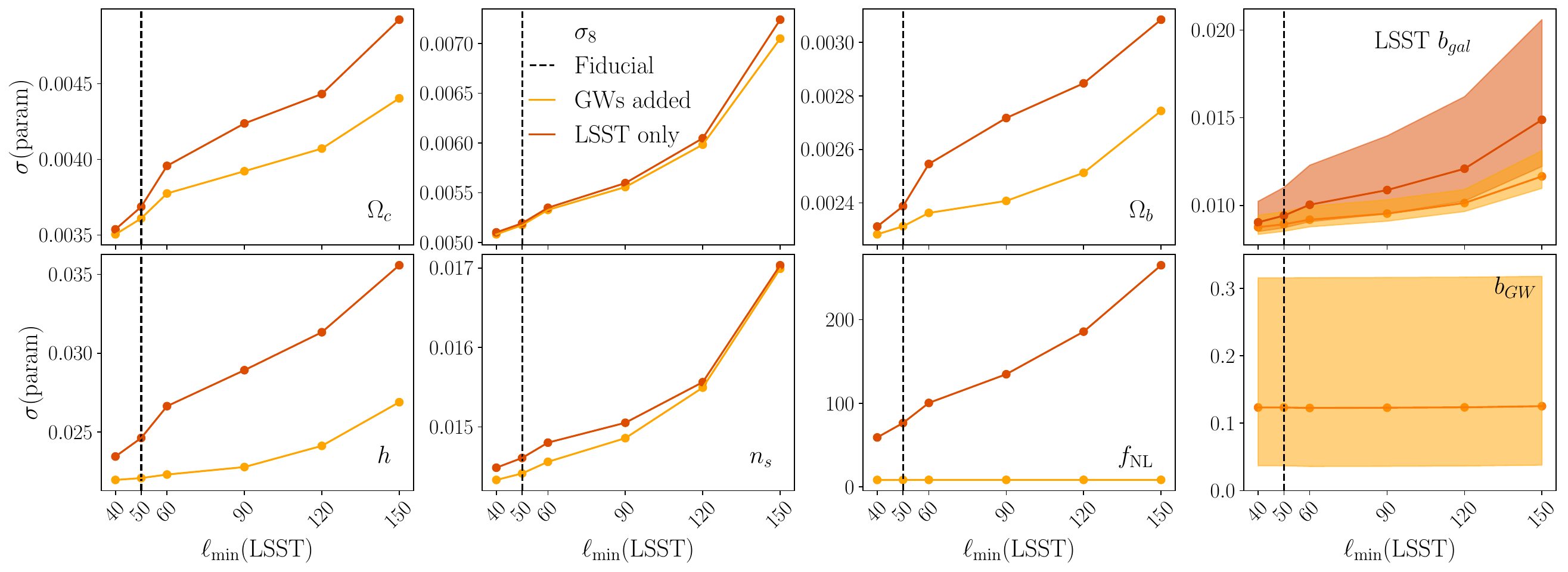}
\end{center}
\caption{Effect of varying $\ell_{\rm min}(\rm LSST)$ on constraining power from LSST 3$\times$2 pt with (yellow) and without (red) GW, for cosmological parameters in $\Lambda$CDM$+\fnl$. For the LSST and GW bias plots, solid lines show the average value and shaded areas show the range of values across redshift bins. The fiducial setup throughout the paper is $\ell_{\rm min}({\rm LSST})=50$, indicated by the vertical dashed line in each column. All analyses use GW Setup 3 ($0<z<3$, $2<\ell<57-90$ with $\ell_{\rm max}$ depending on redshift bin).
}
\label{fig:LSST_ellmin}
\end{figure*}

\begin{figure*}
\begin{center}
\includegraphics[width = \textwidth]{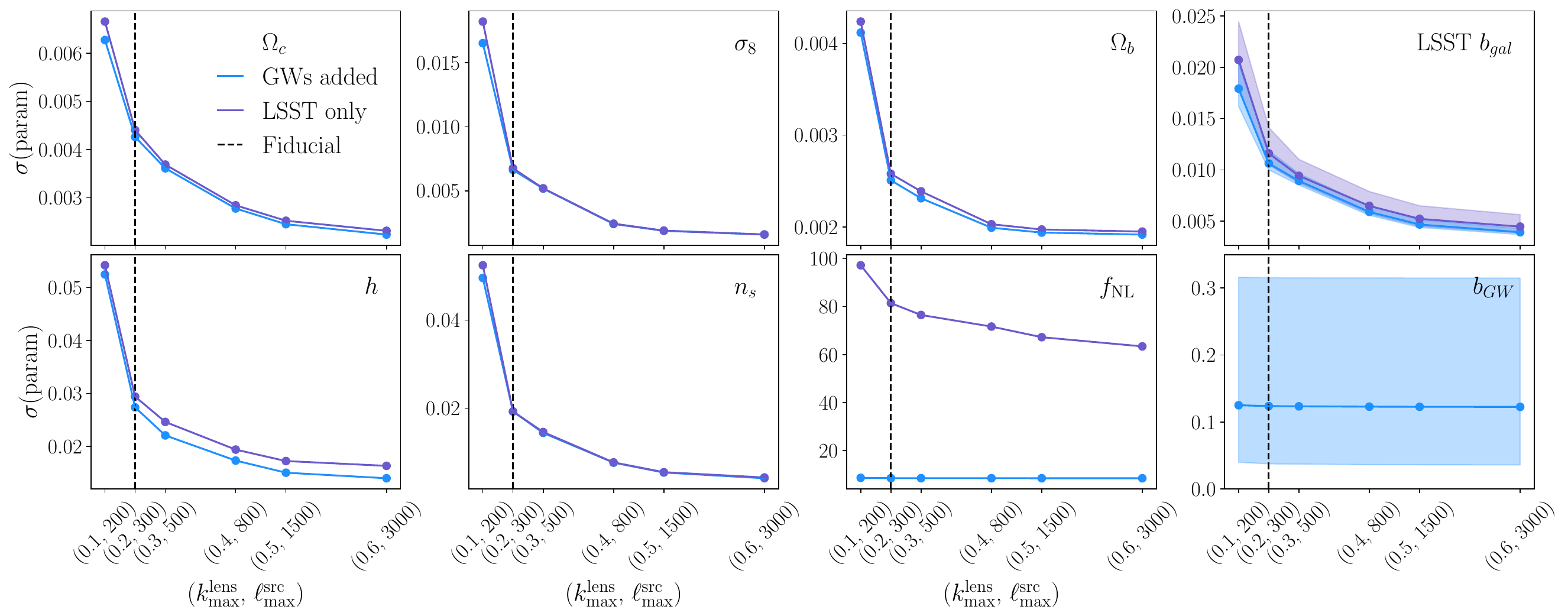}
\end{center}
\caption{Effect of varying $k_{\rm max}^{\rm lens}$ and $\ell_{\rm max}^{\rm src}$ on constraining power for cosmological parameters in $\Lambda$CDM$+\fnl$ for LSST 3$\times$2 pt with (blue) and without (purple) GW. For the LSST and GW bias plots, solid lines show the average value and shaded areas show the range of values across bins. The fiducial values throughout the paper are $k_{\rm max}^{\rm lens}= 0.3h$/Mpc and $\ell_{\rm max}^{\rm src}= 500$. All analyses use GW Setup 3 ($0<z<3$, $2<\ell<57-90$ with $\ell_{\rm max}$ depending on redshift bin).}
\label{fig:kmax}
\end{figure*}

\subsection{Sensitivity of results to analysis choices} \label{subsec:analysis_choices}

We now investigate how changes in the LSST large- and small-scale cuts, the assumed galaxy bias of GW sources, and the GW large-scale cut would affect our results.

\begin{table*}
    \begin{center}
    \textbf{Impact of Varying Analysis Choices}\\
    \begin{tabular}{c c c c c c c c}

    \hline
      & Fiducial & $\ell_{\rm min}^{\rm LSST}$=90 & $k_{\rm max}^{\rm src}$=0.6, $\ell_{\rm max}^{\rm lens}=3000$ & GW bias = $b_{\rm GW1}$ & GW bias = $b_{\rm GW2}$ & $\ell_{\rm min}^{\rm GWs}$=5 & \\ [2pt]
    \hline
    \multicolumn{8}{c}{Relative constraining power: FoM(+GWs)/FoM(LSST)}\\
    $\Omega_c,\sigma_8,\Omega_b,h,n_s$ &
    1.09 & 1.16 & 1.07 & 1.38 & 1.09 & 1.08\\
    $\Omega_c,\sigma_8,\Omega_b,h,n_s,\fnl$ &
    6.95 & 9.13 & 4.70 & 45.3 & 6.33 & 4.72 \\
    $\text{LSST }b_{\rm gal}$ &
    1.01 & 1.01 & 1.01 & 1.06 & 1.01 & 1.01 \\
    \hline
    \multicolumn{8}{c}{Uncertainty on single parameters using LSST+GW} \\
    $\sigma(\fnl)$ &
    8.46 & 8.49 & 8.43 & 1.56 & 9.31 & 14.2 \\
    $\sigma(b_{\rm GW})$, bin 1 &
    0.037 & 0.038 & 0.036 & 0.061 & 0.041 & 0.037 \\
    $\sigma(b_{\rm GW})$, bin 10 &
    0.091 & 0.092 & 0.91 & 0.090 & 0.090 & 0.092 \\
    $\sigma(b_{\rm GW})$, bin 19 &
    0.32 & 0.32 & 0.31 & 0.22 & 0.37 & 0.32 \\

    \end{tabular}
    \caption{Statistics describing the effect on our results of varying analysis choices.  Upper 3 rows: ratio of figure of merit values, \textbf{FoM}($\boldsymbol{p}$)$_{LSST+GW}$/\textbf{FoM}($\boldsymbol{p}$)$_{LSST only}$, for $\boldsymbol{p}= \{\Omega_c,\sigma_8,\Omega_b,h,n_s\}$ ($\Lambda$CDM), then $\boldsymbol{p}=\{\Omega_c,\sigma_8,\Omega_b,h,n_s,\fnl\}$ ($\Lambda$CDM $+\fnl$), and $\boldsymbol{p}$ = [LSST lens galaxy bias values].  Lower 2 rows:  $\sigma_{\text{LSST+GW}}$(\textit{param}) values for $\fnl$ and for a range of GW galaxy biases, for $\Lambda$CDM$+\fnl$.  All statistics use a GW setup with the redshift and $\ell$ range of Setup 3 ($0<z<3$, $2<\ell<57-90$ with $\ell_{\rm max}$ depending on redshift bin.) As expected, raising $\ell_{\rm min}(\rm LSST)$ reduces the total amount of information gathered, and vice versa. Constraints on $\fnl$ already come almost entirely from the GW clustering, so raising $\ell_{\rm min}(\rm LSST)$ has little impact on the $\fnl$ uncertainty after the addition of GWs. We also see that higher bias leads to a higher clustering amplitude and therefore tighter constraints on $\Lambda$CDM parameters from the combination of LSST and GW, when compared to the fiducial case. With higher bias, the amplitude of the scale-dependent bias due to $\fnl$ is larger (as predicted, see Eqn.~\ref{eqn:b_fnl}), leading to significantly stronger constraints on $\fnl$ as well. Also as expected, the loss of the largest scales decreases constraining power on $\fnl$, but we find that even with $\ell_{\rm min}(\rm GWs)=5$, GWs still provide a significant addition to the constraints.
    \label{tab:verification_table}}
    \end{center}
\end{table*}

\subsubsection{LSST scale cuts} \label{subsubsec:ellmin_ellmax_LSST}
Since the maximum angular scale reachable by LSST 3$\times$2pt analysis is not entirely certain, we investigate how different values of $\ell_{\rm min}(\rm LSST)$ would impact our results. We summarize the results in Fig.~\ref{fig:LSST_ellmin}. As expected, raising $\ell_{\rm min}(\rm LSST)$ reduces the total amount of information gathered, and vice versa.
Constraints on $\fnl$ already come almost entirely from the GW clustering, so raising $\ell_{\rm min}(\rm LSST)$ has little impact on the $\fnl$ uncertainty after the addition of GWs.  As an example, constraints using $\ell_{\rm min}(\rm LSST)=90$ are listed in Table~\ref{tab:verification_table}. Note that $\ell_{\rm min}(\rm LSST)$ is determined independently of $\ell_{\rm max}(\rm GW)$, so changes in the former are not implicitly referring to changes in the latter. Thus, the range of scales for LSST are allowed to overlap that of the GWs.

In Fig.~\ref{fig:LSST_ellmin} we see different sensitivities of the $\Lambda$CDM parameters to the LSST large-scale cut.  In particular, as $\ell_{\rm min}(\rm LSST)$ increases, the addition of GW information has greater relative impact upon the constraints on $\Omega_c$, $\Omega_b$, $h$, and $b_{\rm gal}$ (i.e., the gap between LSST only and LSST + GW points increases). This greater dependence on large scales is due in part to the relation of $\Omega_m$ ($=\Omega_c+\Omega_b$) and $h$ to the location of the peak of the 3D power spectrum, $\kpeak$. To see this, we note that the value of $\kpeak$ is close to the comoving value of $k_{\mathrm{eq}}$, the inverse Hubble scale at the time of matter-radiation equality, and $k_{\mathrm{eq}}$ is proportional to $(\Omega_m h^2)^{1/2}$. For the fiducial $\Lambda$CDM parameters, $\kpeak \simeq 0.0187 h$/Mpc, which for the LSST redshift range corresponds to $\sim20<\ell<80$.  Further information on $\Omega_b$ and $\Omega_c$ is also found at higher $k$, but $h$ is particularly dependent on large scales, which is why we see GWs impact constraints most prominently in the $h$ plot of Fig.~\ref{fig:LSST_ellmin}, and in the $h$ contours of Fig.~\ref{fig:fnl_contours}. In contrast, we see that $\sigma_8$ and $n_s$ are much less reliant on large scales--removing LSST information on large scales does not cause the addition of GW large-scale information to become more impactful.

We also test the impact of more aggressive small-scale cuts for the LSST sources and lenses.  Results for $k_{\rm max}^{\rm lens}=0.6 h$/Mpc, $\ell_{\rm max}^{\rm src}=3000$ are listed in Table~\ref{tab:verification_table} (column 4), and results for a range of $(k_{\rm max}^{\rm lens}$, $\ell_{\rm max}^{\rm src})$ values are shown in Fig.~\ref{fig:kmax}.  Little information on $\fnl$ is found at small scales, so while a more aggressive small-scale cut strengthens the constraints on the other $\Lambda$CDM parameters, the impact on $\fnl$ from adding GWs is insensitive to this change.

\subsubsection{Bias of GW-source host galaxies} \label{subsubsec:GW_bias}
Given the relatively small number of GW events to date, the bias for the host galaxies of GW sources is not yet well understood. While the parameterization for $b_{\rm GW}$ adopted in our analysis above is reasonable, it is important to test the impact on our results of a change in the galaxy bias values for the GW sources. We consider two alternative GW galaxy bias models to the fiducial model of Eqn.~\ref{eq:GW_bias}:
\begin{align}
    & b_{\rm GW1} = 1.20(1+z)^{0.59} + 3, \notag \\
    & b_{\rm GW2} = 2.0.  
\end{align}

These alternatives represent 1) a bias model with the same shape as the fiducial bias model but shifted to a much higher value to serve as a conservative upper limit, and 2) a model with a constant (redshift-independent) bias approximately equal to the average of the fiducial values. We have intentionally made extreme choices for the two alternative bias models in order to cover a broad range of possibilities for the true GW bias--redshift relation in our Universe (which is currently unknown).

For model $b_{\rm GW1}$, the higher bias leads to a higher clustering amplitude and therefore tighter constraints on $\Lambda$CDM parameters from the combination of LSST and GW, when compared to the fiducial case. Moreover, in this case, the amplitude of the scale-dependent bias due to $\fnl$ is larger (as expected, see Eqn.~\ref{eqn:b_fnl}), leading to significantly stronger constraints on $\fnl$ as well --for example, in going from the fiducial $b_{\rm GW}$ to $b_{\rm GW1}$, $\sigma(\fnl)$ improves from $8.46$ to $1.56$. These results are shown in Table~\ref{tab:verification_table}, columns 5 and 6.

Table~\ref{tab:verification_table} also shows the constraints after removing the redshift dependence in the GW bias. Following Equation \eqref{eq:GW_bias}, the fiducial bias value in each bin is now set to the average value over all bins. This changes the constraints by less than $10\%$, and highlights the relative insensitivity of our results to choices in the redshift evolution of the fiducial bias values. Note that this insensitivity is expected, as the bias values in each redshift bin are still a free paramenter in our model and is varied jointly with cosmology. Thus our constraints will not be that sensitive to reasonable variations in the fiducial GW bias values --- which only determine the point around which derivatives are estimated --- as is verified in Table~\ref{tab:verification_table}.

\subsubsection{GW scale cuts} \label{subsubsec:ellmin_GW}
While an $\ell_{\rm min}(\rm GWs)$ value of 2 is reasonable since GW detectors can observe across the full sky, we noted before that the varying sensitivity of sky localization, and other factors, across the sky results in an \textit{effective sky fraction} that is below 1. This is particularly relevant when studying the $\fnl$ model considered here, as the signal peaks on large scales (see Section \ref{sec:fnl_into}). We therefore investigate whether the clustering of GW sources still improves constraints from the LSST 3$\times$2pt analysis, if the GW clustering measurements are limited in the largest scales they access. We alter the limit from $\ell_{\rm min}(\rm GWs) = 2 \rightarrow 5$, which mimics a roughly factor of 2 change in the sky area. As expected, the loss of the largest scales decreases constraining power on $\fnl$, but we find that even with $\ell_{\rm min}(\rm GWs)=5$, GWs still provide a significant addition to the constraints (Table~\ref{tab:verification_table}, column 7).


\section{Conclusions} \label{sec:conclusions}

In this work, we have studied the cosmological information that can be extracted from two-point correlation measurements on very large scales, focusing on the clustering of gravitational wave (GW) sources and its combination with two-point measurements from optical galaxy surveys on intermediate to large scales.
The advantages of using GW detectors at the largest scales is that their angular selection function is well understood, they are sensitive to events over the full sky, they can extend to high redshifts, and the distance error per event is relatively small; moreover, they are not subject to the kinds of spatially varying observational and astrophysical systematics that afflict optical surveys on large angular scales. The disadvantages of using GW sources are the relatively low number density of sources (and therefore high shot noise) and poor sky localization (and therefore lack of small-scale information) compared to optical surveys. Given these relative advantages and disadvantages, we have explored the potential benefits of combining GW and optical survey measurements in a complementary way. We investigated the problem in the context of $\Lambda$CDM, $w$CDM, and $\Lambda$CDM$+\fnl$ and considered the combination of several future GW detector setups with projected LSST Y10 $\times$2pt measurements, varying a number of assumptions about the analysis. 

Our main findings are the following:

\begin{itemize}
    \item In $\Lambda$CDM and $w$CDM, there is no significant information gain (improvement in cosmological parameter constraints) from clustering on very large scales ($\ell<50$). 
    However, the LSST+GW combination does enable us to constrain the redshift-dependent bias of the host galaxies of GW sources to an average $\sim$6\% precision, which is a currently largely unknown quantity.
    \item For $\Lambda$CDM$+\fnl$, the addition of very large-scale information, either from LSST 3$\times$2pt or the clustering of GW sources, results in a roughly order of magnitude improvement in the constraint on $\fnl$, with $\sigma(\fnl)=8.8$. For GW to be competitive, the network of detectors must have sensitivity to detect more than $10^5$ events per year out to redshifts up to $z \sim 3$. We also find that including GW analysis of the range $30<\ell<90$ is relatively unimportant, as information on $\fnl$ is found at even larger scales, and LSST already provides sufficient information on the $50<\ell<90$ range in the baseline analysis. 
    \item Changes to the scale cuts in the LSST 3$\times$2pt analysis, to parameterization of the GW bias, and to the scale cuts on the GW analysis do not significantly affect the forecast constraints on the $\Lambda$CDM parameters. However, we find that a change in the GW bias parameterization can change the projected constraints on $\fnl$ significantly, due to the relationship between $\fnl$ and galaxy bias.
\end{itemize}

With the upcoming wealth of data expected from a number of optical/NIR galaxy surveys (LSST, Euclid, Roman) and GW experiments (Einstein Telescope,  Cosmic Explorer and LIGO Voyager), we will soon enter a regime where cleverly combining different datasets could allow us to tackle some of the systematic effects that are challenging to address in a single survey. Here we have investigated how using GW sources as LSS tracers can evade the large-scale systematic effects in optical galaxy surveys, but it is likely that these combinations would have other benefits worth studying, just as there have been many applications of combining galaxy and CMB datasets in the past 10 years \citep{Schaan2017, DES2023_5x2pt}. The combination of galaxy surveys and GW experiments could have a similar potential, opening up a new area of multi-probe cosmology.   


\section*{Acknowledgements}

We would like to especially thank Aditya Vijaykumar for many useful discussions and help with the gravitational waves observations setup. We also thank Daniel Holz, Jose Maria Ezquiaga Bravo, and Hayden Lee for helpful discussions. We thank Fran\c{c}ois Lanusse for help on \texttt{jax-cosmo}. 

ELG is supported by DOE grant DE-AC02-07CH11359. DA is supported by NSF grant No. 2108168. JP is supported by the Eric and Wendy Schmidt AI in Science Postdoctoral Fellowship, a Schmidt Futures program.  CC is supported by the Henry Luce Foundation and DOE grant DE-SC0021949. JF is supported in part by the DOE at Fermilab and in part by the Schmidt Futures program. 


\appendix

\AppendixSection{Covariance Matrix}\label{appx:CovMat}

\begin{figure}
\includegraphics[width = 0.45\textwidth]{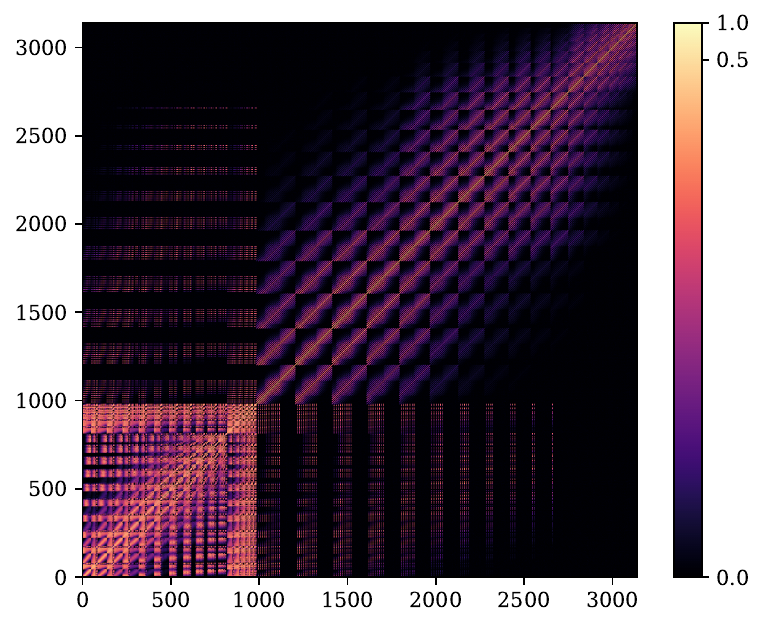}
\caption[width = 0.45\textwidth]{The (Gaussian) correlation matrix of the full data-vector, which includes LSST and the GW sample. The plot uses a log scale for optimized readability. The colors show the correlation strength, which is bounded from below by 0 as our Gaussian covariance model has no anti-correlation terms. The x- and y-axis show the index of the datapoint. There are two visible blocks in this matrix: the lower left block are the LSST bins, and the the upper right block are the GW bins. The off-diagonal ``blocks'' show small amounts of cross-correlation between GW and LSST bins where their $\ell$ ranges overlap.}
\label{fig:cov_appendix}
\end{figure}

Figure \ref{fig:cov_appendix} shows the covariance matrix as described in Equation \eqref{eq:cov}. We show this for the largest datavector of our analysis --- the combination of LSST with ``Setup 3'' of the GW detectors, which includes 15 tomographic bins. The lower left area of the plot represents the LSST bins; we see significant cross-correlation between bins.  The upper right portion of the plot represents the GW bins. On the sides of the plot, we see small amounts of cross-correlation between GW and LSST bins where their $\ell$ ranges overlap.

\bibliography{library}
\bibliographystyle{mnras_2author}



\label{lastpage}
\end{document}